\documentclass{emulateapj}
\usepackage{natbib}
\usepackage{gensymb}
\usepackage{xcolor}
\usepackage[normalem]{ulem}

\newcommand\etal{et al.}
\newcommand\ie{i.e.}
\newcommand\eg{e.g.}

\newcommand\kms{\ifmmode{\rm km\ s^{-1}}\else$\rm km\ s^{-1}$\fi}
\newcommand\ergs{\ifmmode{\rm erg\ s^{-1}}\else$\rm erg\ s^{-1}$\fi}

\def\eps@scaling{1.0}
\newcommand\plotthree[3]{{
 \typeout{Plotthree included the files #1 #2 #3}
 \centering
 \leavevmode
 \columnwidth=.33\columnwidth
 \includegraphics[width={\eps@scaling\columnwidth}]{#1}
 \includegraphics[width={\eps@scaling\columnwidth}]{#2}
 \includegraphics[width={\eps@scaling\columnwidth}]{#3}
}}
\newcommand\plotrtwo[2]{{
 \typeout{Plotrtwo included the files #1 #2}
 \centering
 \leavevmode
 \columnwidth=.5\columnwidth
 \includegraphics[angle=270, width={\eps@scaling\columnwidth}]{#1}
 \includegraphics[angle=270, width={\eps@scaling\columnwidth}]{#2}
}}
\shorttitle{N5813 FROM CHANDRA}
\shortauthors{RANDALL ET AL.}
\slugcomment{Accepted for publication in ApJ, v. \today}

\begin{document}

\title{A Very Deep Chandra Observation of the Galaxy Group
  NGC~5813: AGN Shocks, Feedback, and Outburst History} 

\author{S. W. Randall\altaffilmark{1},
  P. E. J. Nulsen\altaffilmark{1}, C. Jones\altaffilmark{1},
  W. R. Forman\altaffilmark{1}, E. Bulbul\altaffilmark{1},
  T. E. Clarke\altaffilmark{2}, 
  R. Kraft\altaffilmark{1}, E. L. Blanton\altaffilmark{3}, 
  L. David\altaffilmark{1}, N. Werner\altaffilmark{4},
  M. Sun\altaffilmark{5}, M. Donahue\altaffilmark{6},
  S. Giacintucci\altaffilmark{7}, A. Simionescu\altaffilmark{8}}

\altaffiltext{1}{Harvard-Smithsonian Center for Astrophysics, 60
  Garden St., Cambridge, MA 02138, USA; srandall@cfa.harvard.edu}
\altaffiltext{2}{Naval Research Laboratory, 4555 Overlook Avenue SW,
  Code 7213, Washington, DC 20375, USA} 
\altaffiltext{3}{Astronomy Department and Institute for Astrophysical
  Research, Boston University, 725 Commonwealth Avenue, Boston, MA
  02215, USA}
\altaffiltext{4}{Kavli Institute for Particle Astrophysics and
  Cosmology, Stanford University, 452 Lomita Mall, Stanford,
  California 94305-4085, USA}
\altaffiltext{5}{Department of Physics, University of Alabama in
  Huntsville, Huntsville, AL 35899, USA} 
\altaffiltext{6}{Physics and Astronomy Department, Michigan State
  University, East Lansing, MI 48824, USA}
\altaffiltext{7}{Department of Astronomy, University of Maryland,
  College Park, MD 20742, USA}
\altaffiltext{8}{Japan Aerospace Exploration Agency, 3-1-1 Yoshinodai,
  Sagamihara, Kanagawa 229-8510, Japan}

\begin{abstract}

  We present results from a very deep (650~ks) {\it Chandra} X-ray
  observation of the galaxy group NGC~5813, the deepest {\it Chandra}
  observation of a galaxy group to date.  Earlier observations showed
  two pairs of cavities distributed roughly collinearly, with each
  pair associated with an elliptical shock front.  The new
  observations confirm a third pair of outer cavities, collinear with
  the other pairs, and reveal an associated outer outburst shock, with
  measured temperature jump, at $\sim30$~kpc.  This system is
  therefore unique in exhibiting three cavity pairs, each associated
  with an unambiguous shock front, arising from three distinct
  outbursts of the central AGN.  As such, it is particularly
  well-suited to the study of ongoing AGN feedback.  The implied mean
  kinetic power is roughly the same for each outburst, demonstrating
  that the average AGN kinetic luminosity can remain stable over long
  timescales ($\sim50$~Myr).  The two older outbursts have larger,
  roughly equal total energies as compared with the youngest outburst,
  implying that the youngest outburst is ongoing.  We find that the
  radiative cooling rate and the mean shock heating rate of the gas are well
  balanced at each shock front, suggesting that AGN outburst shock
  heating alone is sufficient to offset cooling and establish AGN/ICM
  feedback within at least the central 30~kpc.  This heating takes
  place roughly isotropically and most strongly at small radii, as is
  required for feedback to operate.  We suggest that shock heating may
  play a significant role in AGN feedback at smaller radii in other
  systems, where weak shocks are more difficult to detect.  We find
  non-zero shock front widths that are too large to be explained by
  particle diffusion.  Instead, all measured widths are consistent
  with shock broadening due to propagation through a turbulent ICM
  with a mean turbulent speed of $\sim$70~\kms.  Finally, we place
  lower limits on the temperature of any volume-filling thermal gas
  within the cavities that would balance the internal cavity pressure
  with the external ICM.  The most stringent limit we find is $kT >
  16$~keV. 

\end{abstract}

\section{Introduction} \label{sec:intro}

Early {\it Chandra} and {\it XMM-Newton} X-ray observations revealed
that the amount of gas cooling to very low temperatures at the centers
of cool core clusters is much less than what is expected from simple
radiative cooling models (David \etal\ 2001; Peterson \etal\ 2001;
Peterson \& Fabian 2006).  The implication is that the diffuse X-ray
emitting gas must be heated, either by pre-heating during cluster
formation or by ongoing energy injection.  The most likely heating
mechanism is feedback due to energy injection by the central active
galactic nucleus (AGN) of the cD galaxy (see McNamara \& Nulsen 2007
and Fabian 2012 for recent reviews).  During this process, matter is
accreted onto the central supermassive black hole (SMBH), which
drives powerful jets.  These jets evacuate cavities in the
intracluster medium (ICM), which
can drive shocks as they are inflated and subsequently rise buoyantly
(Churazov \etal\ 2001).  The energy contained in cavities and shocks
is then available to heat the ICM, which lowers the cooling rate of
the gas and
subsequently the SMBH accretion rate.  The ensuing decrease in AGN
heating allows the gas to once again cool and accrete onto the SMBH,
establishing a feedback loop that regulates the temperature of the
ICM.  Several studies have shown that, generally, the total enthalpy
in cavities in cool core systems is sufficient to offset radiative
cooling in individual galaxies, galaxy groups, and clusters
(B\^{i}rzan \etal\ 2004; Rafferty \etal\ 2006; Nulsen \etal\ 2007;
Hlavacek-Larrondo \etal\ 2012).  However, the details of how and where
this energy gets transferred to the ICM are unclear.  Weak AGN outburst
shocks are also expected to heat the ICM, although they are difficult
to detect and clear examples are very rare.

Galaxy groups provide an excellent opportunity to study AGN feedback.
Their lower temperatures are more easily measured with modern high
angular resolution X-ray satellites, and their central AGN can more
easily disturb the diffuse gas due to their shallower gravitational
potentials as compared with clusters.  
Here we report on results from
a very deep {\it Chandra} observation of the central galaxy in the
galaxy group NGC~5813 (N5813).  This group is a relatively isolated
sub-group in the NGC~5846 group (Mahdavi \etal\ 2005; Machacek \etal\ 2011).
In Randall \etal\ (2011, hereafter R11) we presented results based on an
initial 150~ks {\it Chandra} observation of N5813.
The ICM in this group has a remarkably regular morphology, with
three pairs of roughly collinear cavities and associated surface
brightness edges, and shows no
clear signs of a recent merger event.
(Note that we refer to the diffuse gas in this group as the ICM rather
than the intragroup medium (IGM) throughout to stress the connection
with feedback in clusters and to avoid confusion with the intergalactic medium.)
With clear, cleanly separated signatures from three distinct
outbursts of the central AGN and no other significant dynamical
processes at work, N5813 is uniquely well-suited to the study of AGN
feedback.  In this work, we focus on the implications for AGN feedback
and the outburst history of the central SMBH.

We assume an angular diameter distance to N5813 of 32.2~Mpc (Tonry
\etal\ 2001), which gives a scale of 0.15~kpc/\arcsec.
All uncertainty ranges are 68\% confidence intervals (\ie, 1$\sigma$), unless
otherwise stated.

\section{{\it Chandra} Data Analysis} \label{sec:analysis}

\subsection{Observations and Data Reduction} \label{sec:obs}

The {\it Chandra} observations that were used in the analysis we
present here are summarized in Table~\ref{tab:obs}.  The aimpoint was
on the back-side illuminated ACIS-S3 CCD for each observation.
 All data were reprocessed from the level 1 event files using {\sc
   CIAO} and {\sc CALDB~4.4.3}. CTI and time-dependent gain corrections were
applied. {\sc lc\_clean} was used to remove background
flares\footnote{\url{http://asc.harvard.edu/contrib/maxim/acisbg/}}.
The mean event rate was calculated from a source free region using
time bins within 3$\sigma$ of the
overall mean, and bins outside a factor of 1.2 of this mean were
discarded.  There were no periods of strong background
flares.  The cleaned exposure times are given in Table~\ref{tab:obs},
for a total time of 635~ks.

Diffuse emission from N5813 fills the image FOV for each
observation.  We therefore used the {\sc
  CALDB\footnote{\url{http://cxc.harvard.edu/caldb/}}} 
blank sky background files appropriate for each observation,
normalized to match the 10-12~keV count rate in our observations to
account for variations in the particle background.
To generate exposure maps, we used a MEKAL model with $kT = 0.7$~keV,
Galactic absorption, and abundance of 30\% solar at a redshift $z =
0.006578$.

\section{Image Analysis} \label{sec:image}

\begin{figure}
\plotone{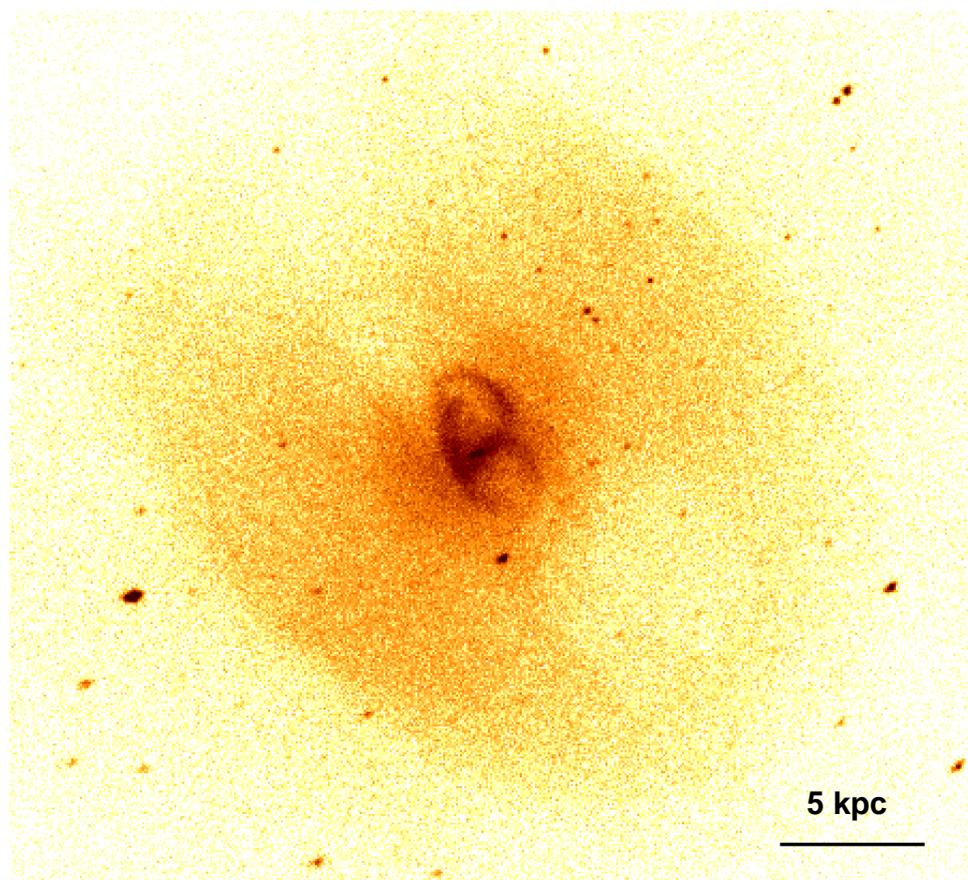}
\caption{
  Exposure corrected, background subtracted, 0.3--3~keV {\it
    Chandra} image of the central region of N5813, 
  unsmoothed and with point sources included (1 pixel = 0.5\arcsec).
\label{fig:core_ximg}
}
\end{figure}

\begin{figure}
\plottwo{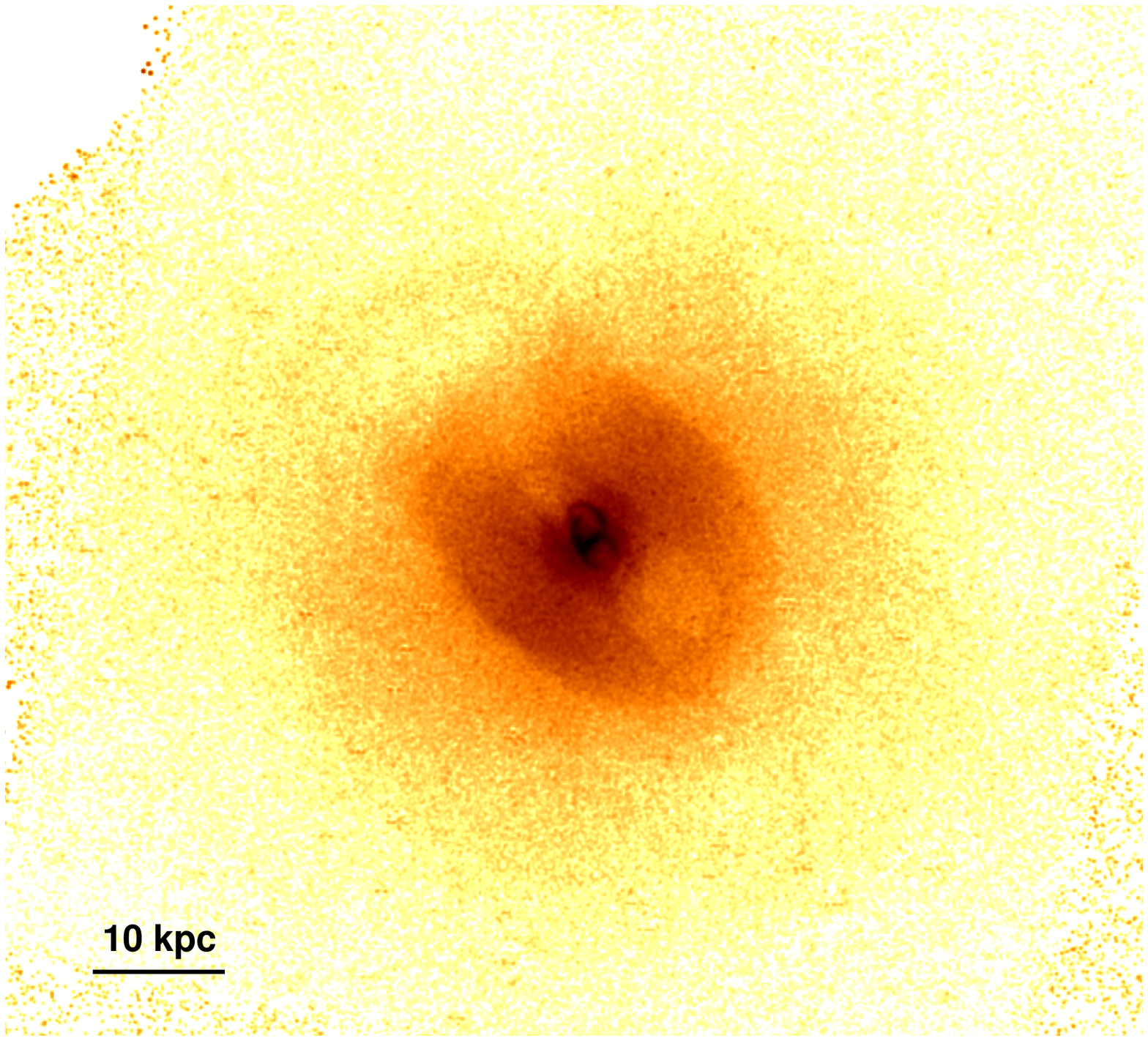}{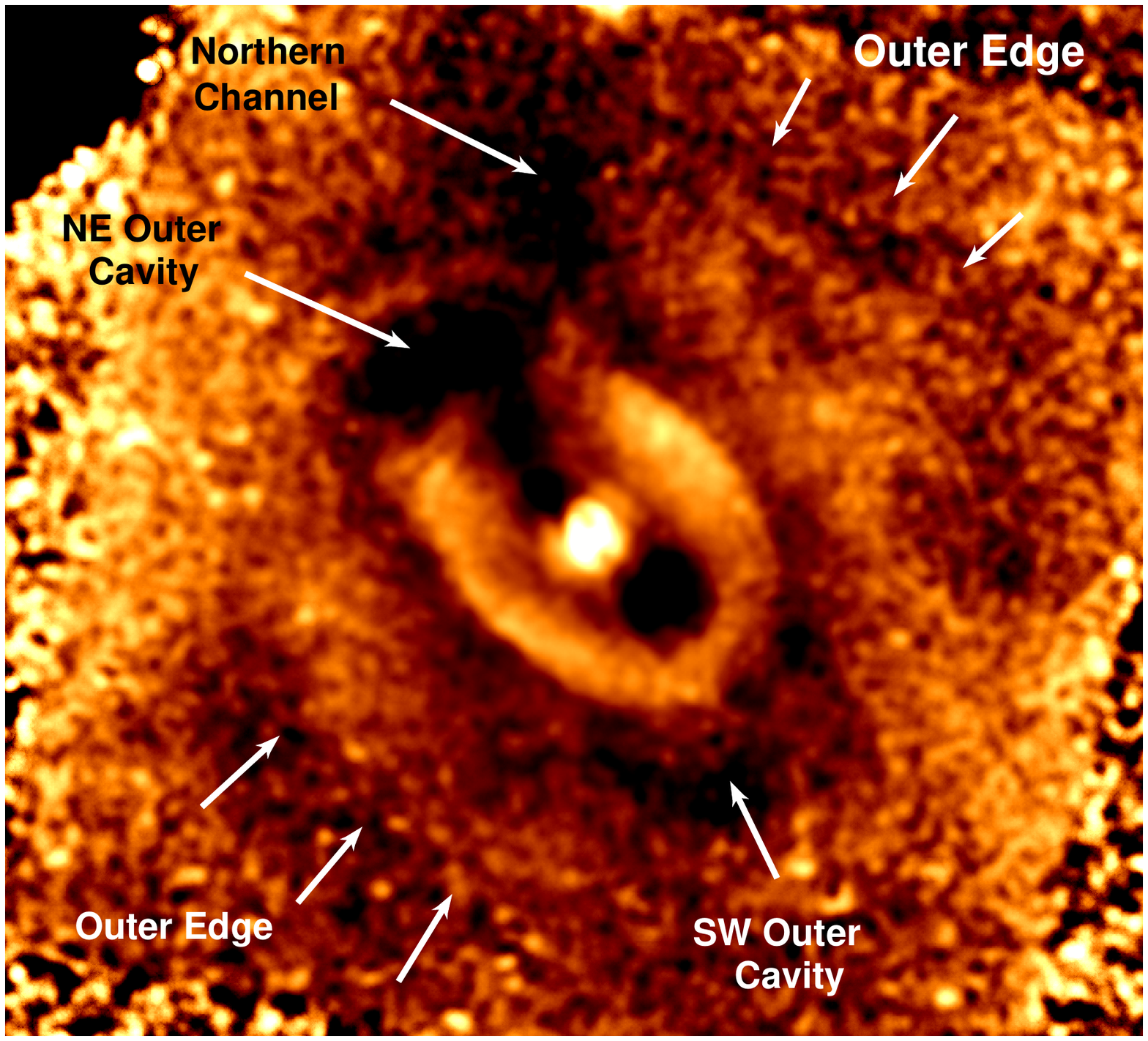}
\caption{
  Left: Exposure corrected, background subtracted, 0.3--3~keV {\it
    Chandra} image, with point sources removed and smoothed with a
  $\sigma=1.5$\arcsec\ Gaussian.  The image shows bright rims
  surrounding an inner pair of cavities, a prominent elliptical edge
  surrounding a pair of cavities at intermediate radii (with the more
  obvious cavity to the SW and the NE cavity apparently broken into
  two connected cavities), and a subtle outer edge associated with a
  faint pair of outer cavities (with the more obvious cavity to the
  NE).  Right: X-ray image divided by a 2D fitted beta model and
  smoothed with a $\sigma=6$\arcsec\ Gaussian, shown on
  the same scale.  The
  outer cavities and edges are more clearly seen in this residual
  image, while the inner cavities are not visible due to the larger
  smoothing scale and saturation of the color scale.  The image also
  reveals a faint ``channel'' of decreased 
  surface brightness extending to the north, apparently connected to
  the NE outer cavity.
\label{fig:ximg}
}
\end{figure}

The exposure corrected, background subtracted, 0.3--3~keV {\it
  Chandra} image of the central region is shown in
Figure~\ref{fig:core_ximg}. To enhance the visibility of the diffuse
emission bright point sources were removed, and the regions containing
point sources were ``filled in'' using a Poisson distribution whose
mean was equal to that of a local annular background region.  To
better show faint structure, particularly in the outer regions, we
fitted the X-ray image with a 2D beta-model using the software package
{\sc sherpa} (Freeman \etal\ 2001) and divided the image by the model
to produce a residual image.  
The smoothed, point source free image and the residual image are shown
in Figure~\ref{fig:ximg}.

  These images clearly show three pairs of roughly collinear cavities,
  along an axis from the SW to the NE.  The inner cavity pair is
  surrounded by bright rims (at about 1~kpc), and each of the
  intermediate and outer 
  cavity pairs is associated with an elliptical surface brightness
  edge (at $\sim$10~kpc and $\sim$30~kpc, respectively).  In R11, we
  showed that the 1~kpc and 10~kpc edges are shock fronts that were
  driven during the expansion phase of their associated cavities.
  The new, deeper observations clearly show a third (outer) cavity pair and
  edge, only hinted at in earlier observations, and allow us to unambiguously
  identify the outer edge as an associated shock front
  (\S~\ref{sec:shprofs}).
  These outer features are most clearly seen in the residual image
  (Figure~\ref{fig:ximg}). Thus, these deep X-ray observations show
  clear signatures from three distinct outbursts of the central AGN.
  We note that, here and throughout, by ``outburst'' we refer to
    the creation of a cavity pair and its associated shocks, rather
    than a rapid increase in AGN jet power, since these features are
    in principle consistent with either a constant or variable jet
    power (see \S~\ref{sec:history}).

  To look for structure in the faint emission at large radii, beyond
  the FOV of {\it Chandra}, we examined the available archival {\it
    XMM-Newton} observations of N5813.  N5813 was observed with {\it
    XMM-Newton} on three occasions: July 23 2005 (ObsID 0302460101),
  February 11 2009 (ObsID 0554680201), and February 17 2009 (ObsID
  0554680301). The event files were calibrated using the XMM-Newton
  Science Analysis System (SAS) version 13.5.0, and the most recent
  calibration files as of 2014 July. The calibrated, cleaned event
  files were produced after filtering for particle background flares.
  This resulted in cleaned exposure times of 120~ks, 127~ks, and 83~ks
  for the MOS1, MOS2, and PN detectors, respectively. Details of the
  data reduction are described in Bulbul \etal\ (2012a).

\begin{figure}
\plottwo{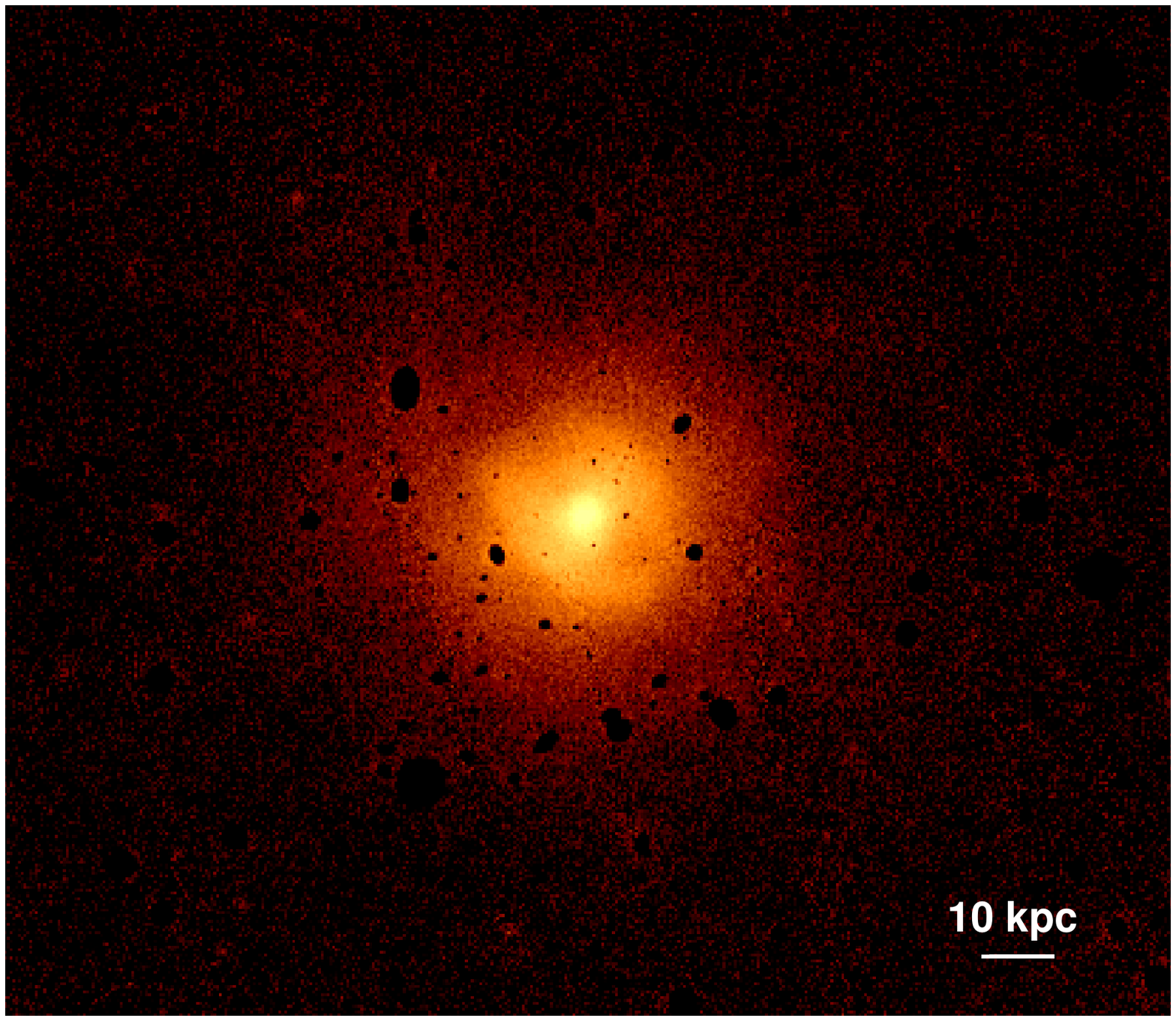}{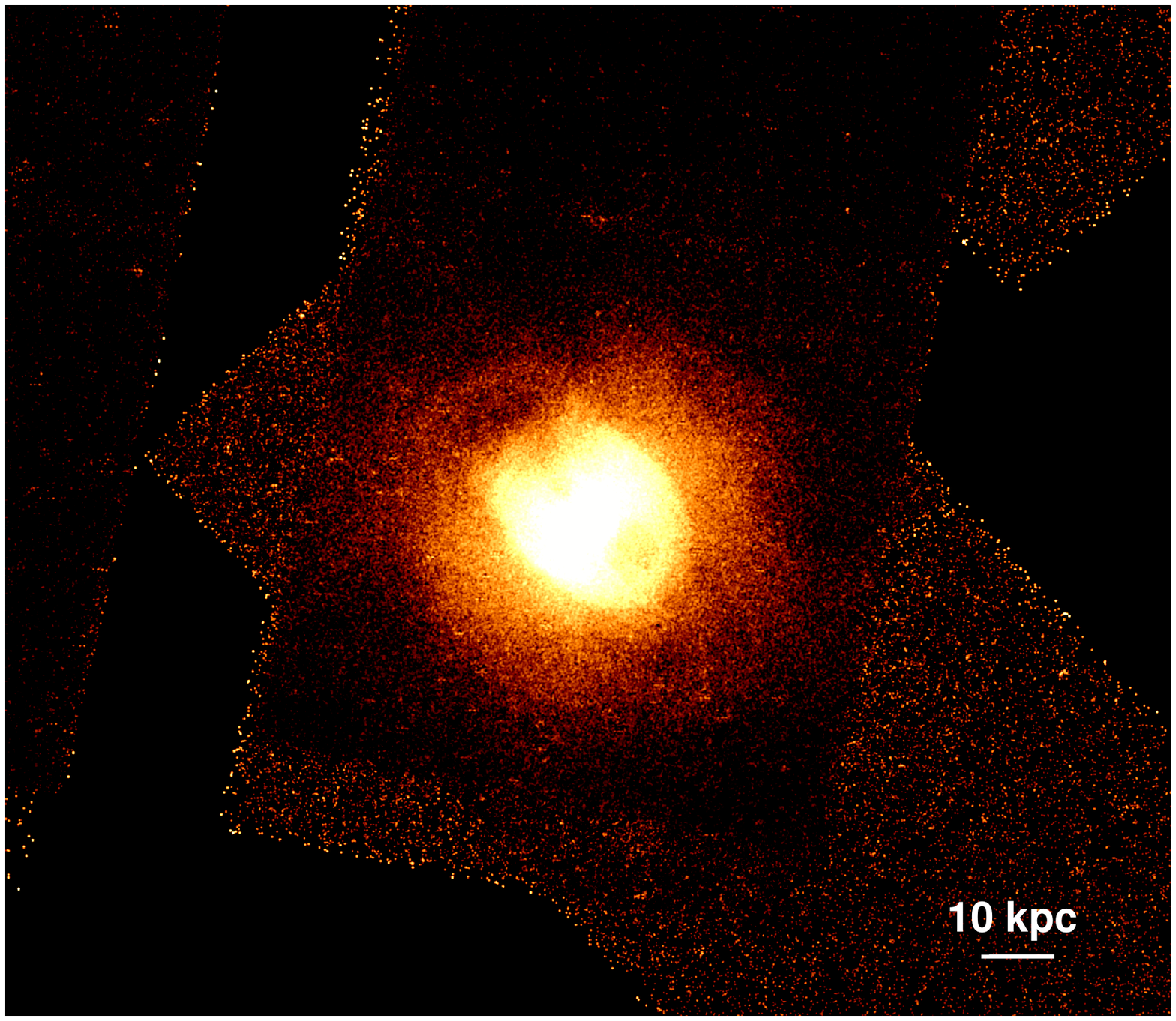}
\caption{
Left: Background and exposure corrected 0.4--7.2~keV {\it XMM-Newton}
image of N5813.  Right: Smoothed {\it Chandra} image shown on the
same scale, with the intensity scale chosen to better show the faint,
outer emission.
\label{fig:xmm_compare}
}
\end{figure}

\begin{figure}
\plotone{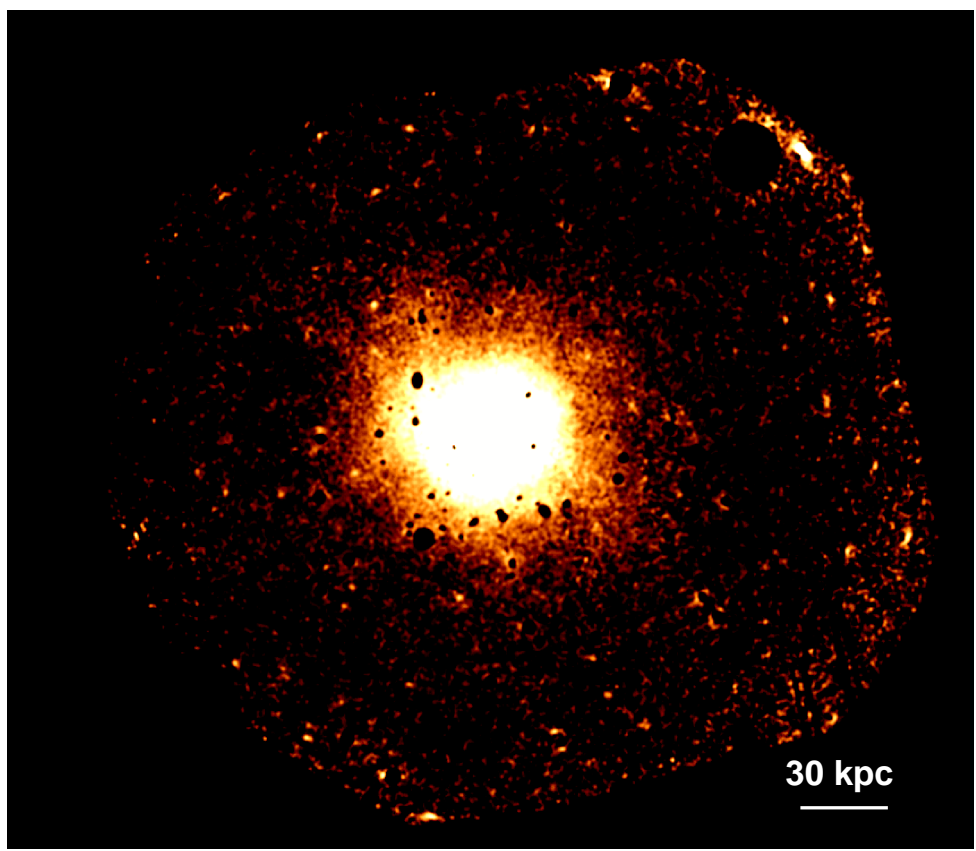}
\caption{
  {\it XMM-Newton} image shown in Figure~\ref{fig:xmm_compare},
  smoothed and with the intensity scale chosen to show the faint
  emission at large radii beyond the {\it Chandra} FOV.  There are no
  additional cavities or shock edges visible beyond the inner features
  identified in the {\it Chandra} images.
\label{fig:xmm_smo}
}
\end{figure}

  The background subtracted, exposure corrected, merged {\it
    XMM-Newton} image is shown in Figure~\ref{fig:xmm_compare} beside
  the {\it Chandra} image.  The 10~kpc shock edges are visible, as are
  the intermediate cavities and the NE outer cavity.  The inner
  $\sim1$~kpc cavities and rims are not resolved.  A smoothed version
  of this image, with the intensity scale chosen to better show faint
  emission at large radii, is shown in Figure~\ref{fig:xmm_smo}.  The
  extended emission is roughly azimuthally symmetric, and shows no
  clear evidence of a fourth pair of cavities or shock edge beyond the
  30~kpc edge detected with {\it Chandra}.  Since, for the purposes of
  this study, we are most interested in the detailed structure in the
  ICM, which is best resolved with {\it Chandra}, we will not consider
  the {\it XMM-Newton} data further here.

\section{Spatially Resolved Spectroscopy} \label{sec:spec}

Unless otherwise specified, all spectra were fitted in the
0.6--3.0~keV band using {\sc xspec}, with an absorbed {\sc apec} model
with the absorption fixed at the Galactic value of $N_H = 4.37\times
10^{20}$~cm$^{-2}$ (Kalberla \etal\ 2005) and the abundance allowed to
vary.  Spectra were grouped with a minimum of 40 counts per energy
bin.  Anders \& Grevesse (1989) abundance ratios and {\sc
  AtomDB~2.0.2} were used throughout.  We note that using this version
of {\sc AtomDB} gives systematically larger temperatures (by roughly
0.05--0.1~keV, significantly larger than our typical statistical
error) as compared with {\sc AtomDB~1.3.1}, which was used in R11.
This difference arises from significant changes in the ionization
balance, particularly at the low ($\sim0.7$~keV) temperatures we
consider here (see Fig.~3 in Foster \etal\ 2012).  Thus, we expect a
systematic offset between the spectroscopic quantities that we derive
here as compared with R11.

\subsection{Spectroscopic Maps} \label{sec:maps}

To study the thermal structure of the ICM, we generated a smoothed
temperature map using the method described in Randall \etal\ (2008),
and also used in R11.  For each pixel in the temperature map, spectra were
extracted from a surrounding circular region that contained 1500 source
counts.  The temperature map pixel value was then set equal to the
best fitting temperature from fitting to these spectra.  Since the
extraction regions are generally larger than the pixels,
nearby pixels are correlated with one another, such that the
temperature map is effectively smoothed on the scale of the local
extraction region size.  Some advantages of this method are that there
is no a priori assumption about the thermal structure of the gas, and
experience has shown that such maps can resolve real structure on
scales that are smaller than the size of the extraction regions.

For comparison, we also constructed temperature maps using the contour
binning method developed by Sanders (2006).  Here, extraction regions
are defined based on surface brightness contours.  Regions were
defined to achieve a signal-to-noise ratio of 38 (corresponding to
$\sim$1,400 counts in bright regions) in each region.  This method
assumes that the temperature distribution follows the surface
brightness distribution (which is frequently, but not always, the
case).  Some advantages of this method are that each extraction region
is independent of the others (as long as the region widths are large
compared with the local PSF), and it is relatively cheap
computationally (as compared with the smoothed temperature map
method).  A comparison of the smoothed and contour binned temperature
maps gives an indication of the extraction region size as a function
of position in the smoothed map.

\begin{figure}
\plottwo{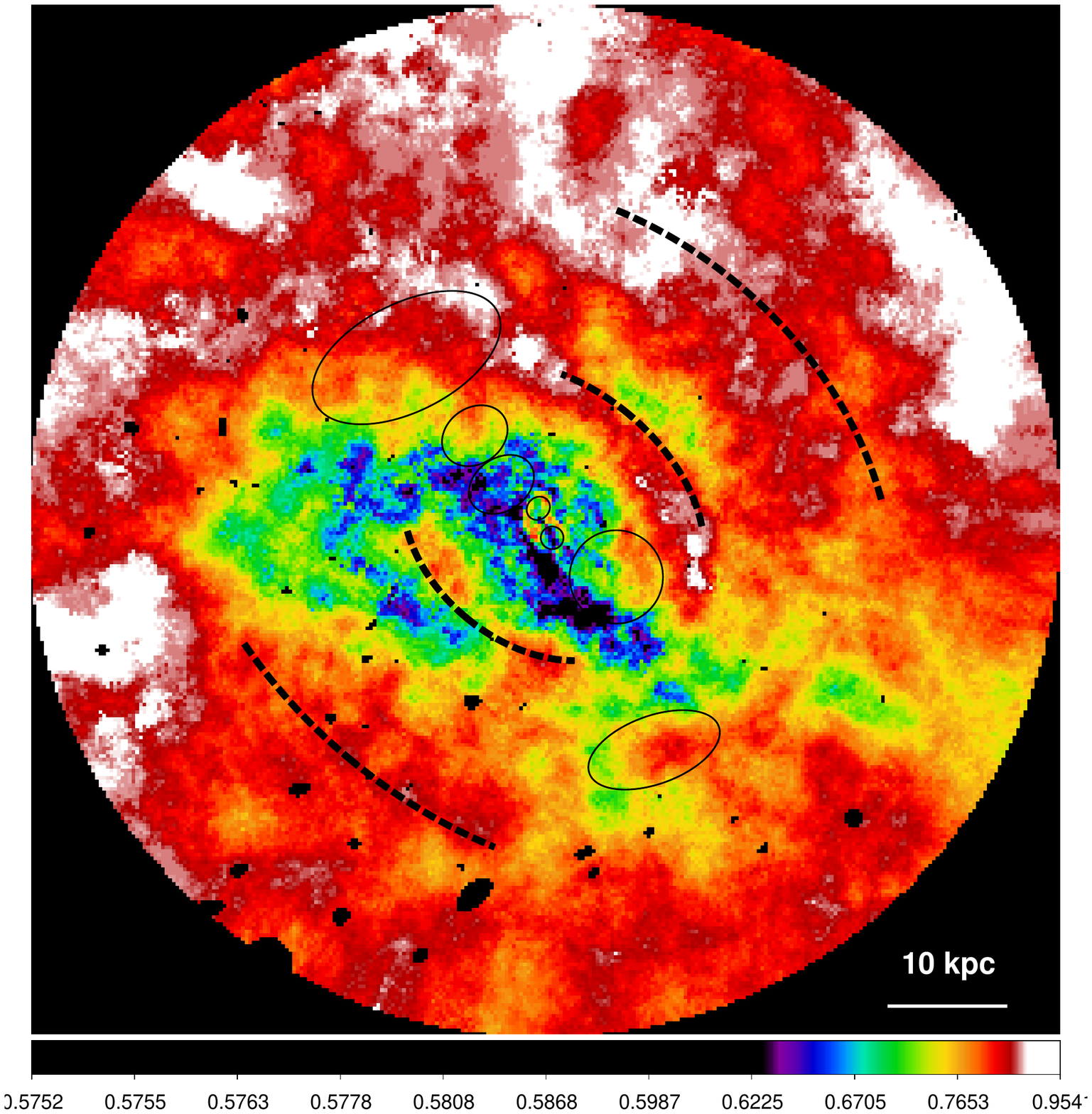}{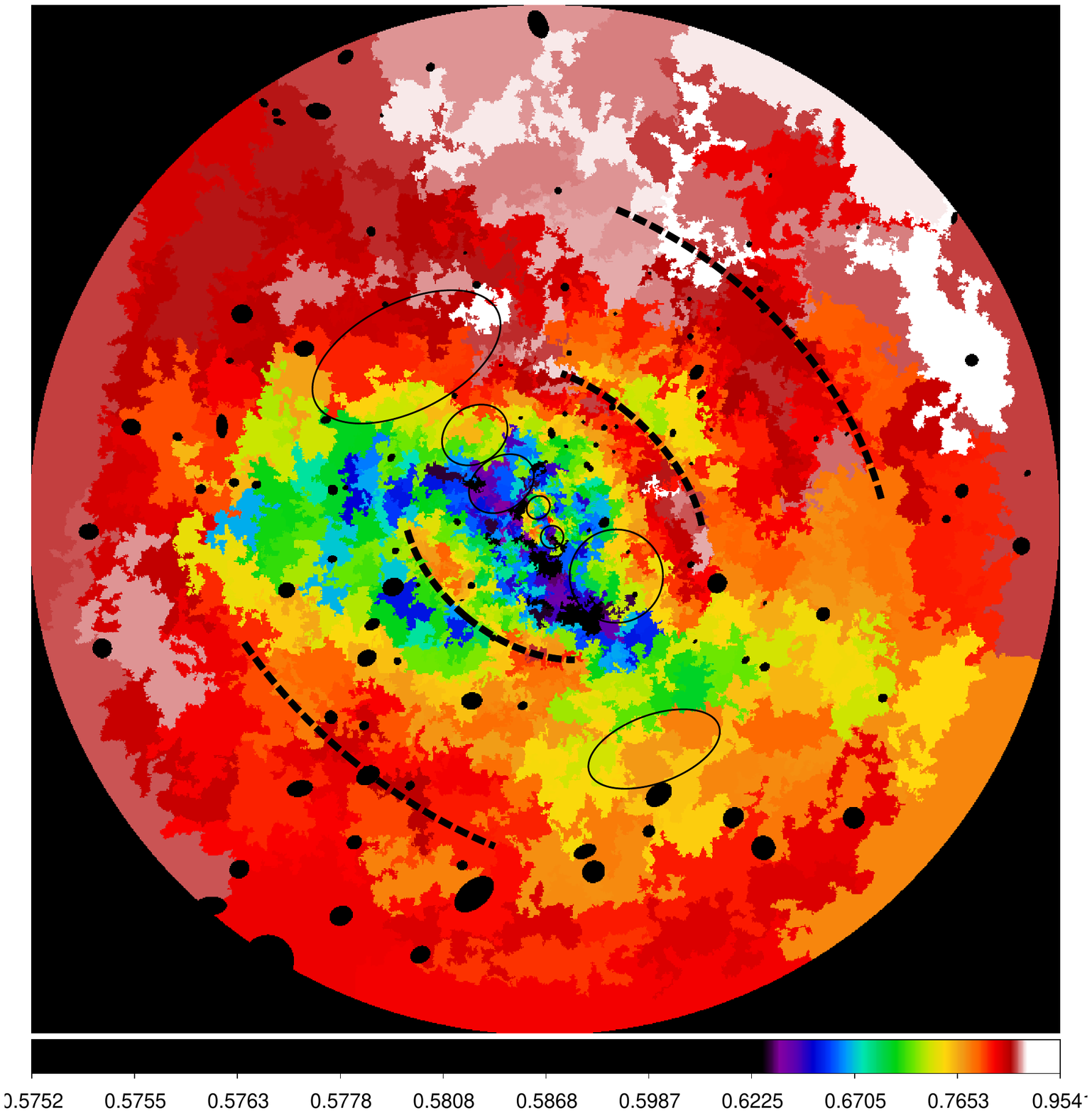}
\caption{
  Smoothed (left) and contour binned (right) temperature maps.  The
  locations of the 10~kpc and 30~kpc shock fronts are indicated with
  dashed black lines, and the cavity locations with black ellipses.
  Both maps clearly show temperature increases associated with the
  1~kpc and 10~kpc shocks, and hint at increases associated with the
  30~kpc shock (particularly to the NW).  Also visible is a SW to NE
  plume of cool gas that has been uplifted by the buoyantly rising
  cavities.  
\label{fig:tmap}
}
\end{figure}

\begin{figure}
\plottwo{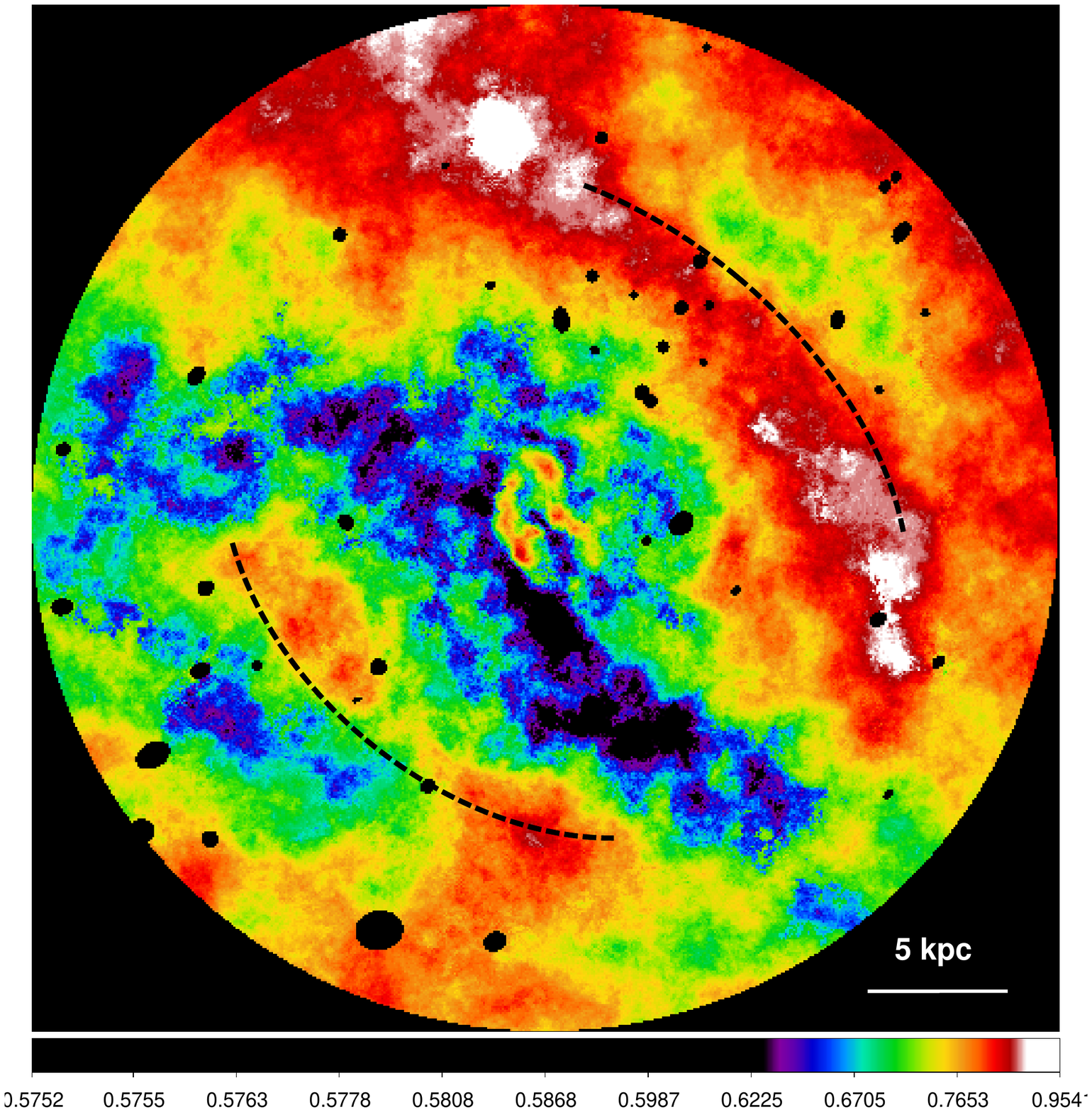}{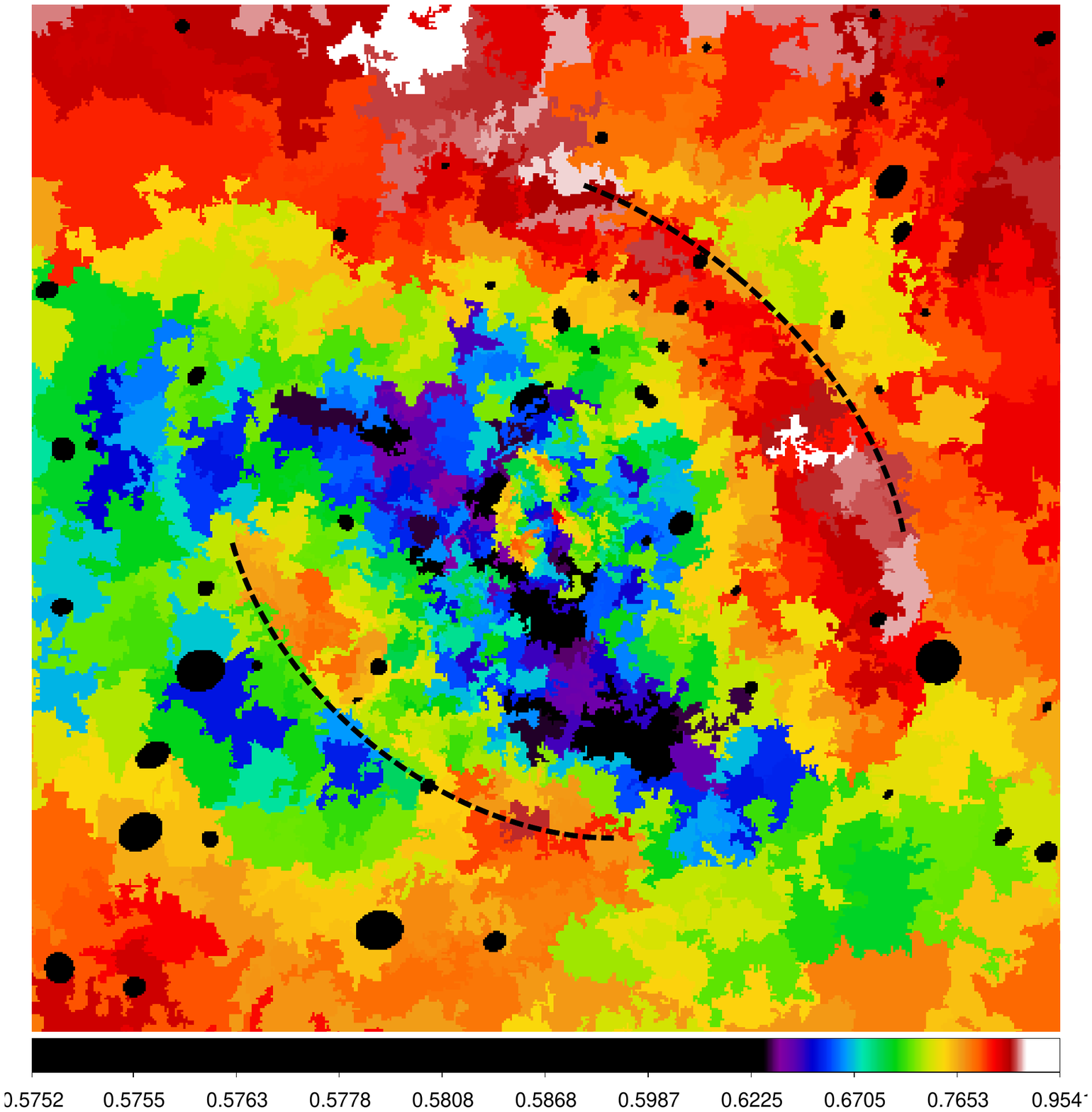}
\caption{
  Left: High-resolution smoothed temperature map of the core region.
  Right: The contour binning temperature map shown on the same scale.
  The 10~kpc shock fronts are indicated as in Figure~\ref{fig:tmap}.
\label{fig:core_tmap}
}
\end{figure}

Smoothed temperature maps over a wider area and a higher resolution
map of the core are shown beside the contour binned maps in
Figures~\ref{fig:tmap}~and~\ref{fig:core_tmap}, respectively.
Both the smoothed and contour binned temperature maps reveal
relatively hot emission associated with the bright rims surrounding
the inner cavities and just inside the prominent 10~kpc surface
brightness edges.  These features were identified as shocks in R11.
The maps suggest additional temperature rises associated with the
outer 30~kpc surface brightness edges, particularly to the NW.  The
temperature profiles across these edges are presented in
\S~\ref{sec:shprofs}.  The temperature maps also show a plume of cooler
gas, roughly extending along the line defined by the cavities, from SW
to NE.  In R11, we showed that this plume is consistent with arising
from cool gas that has been lifted by the buoyantly rising X-ray cavities.

\begin{figure}
\plottwo{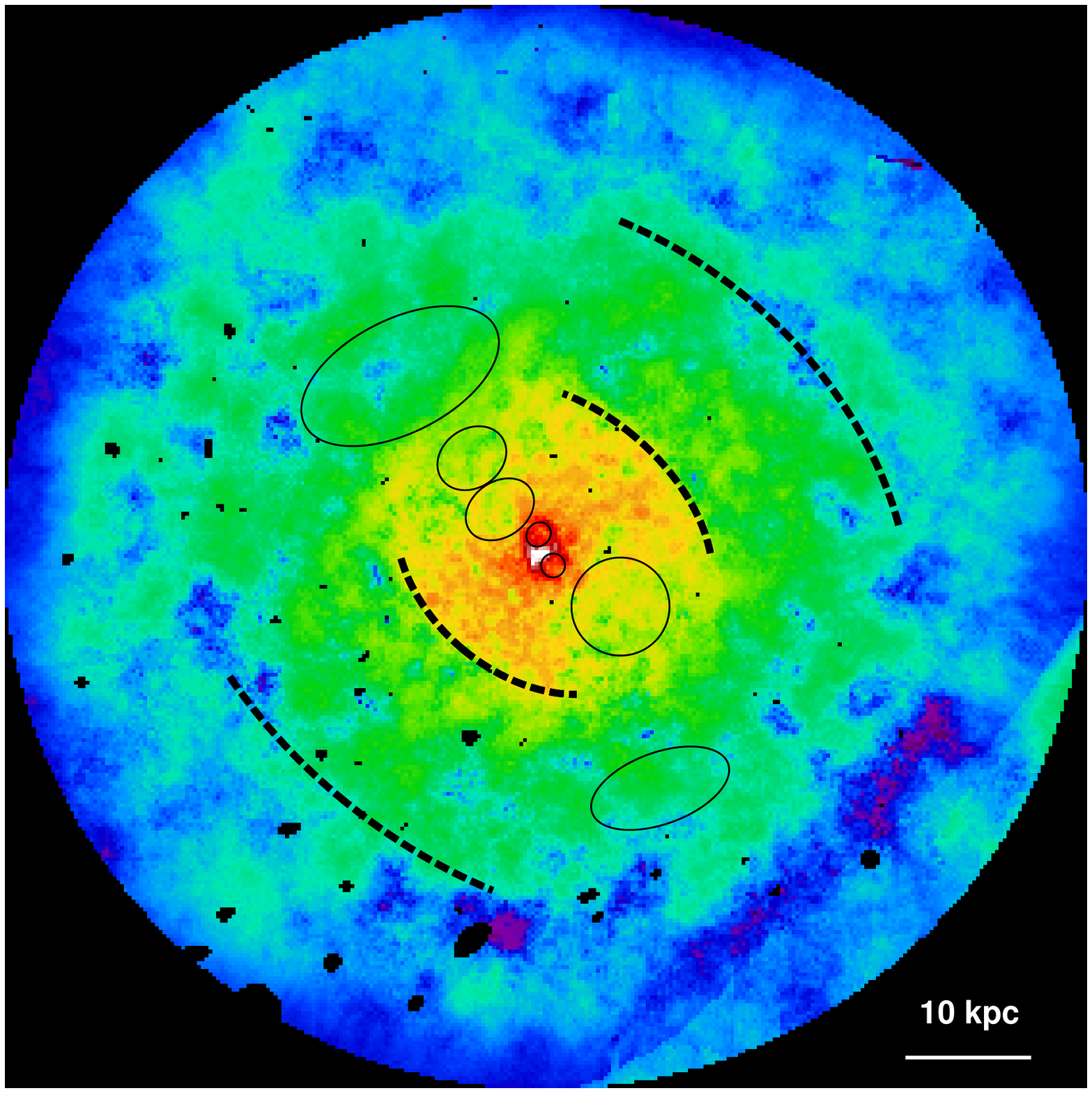}{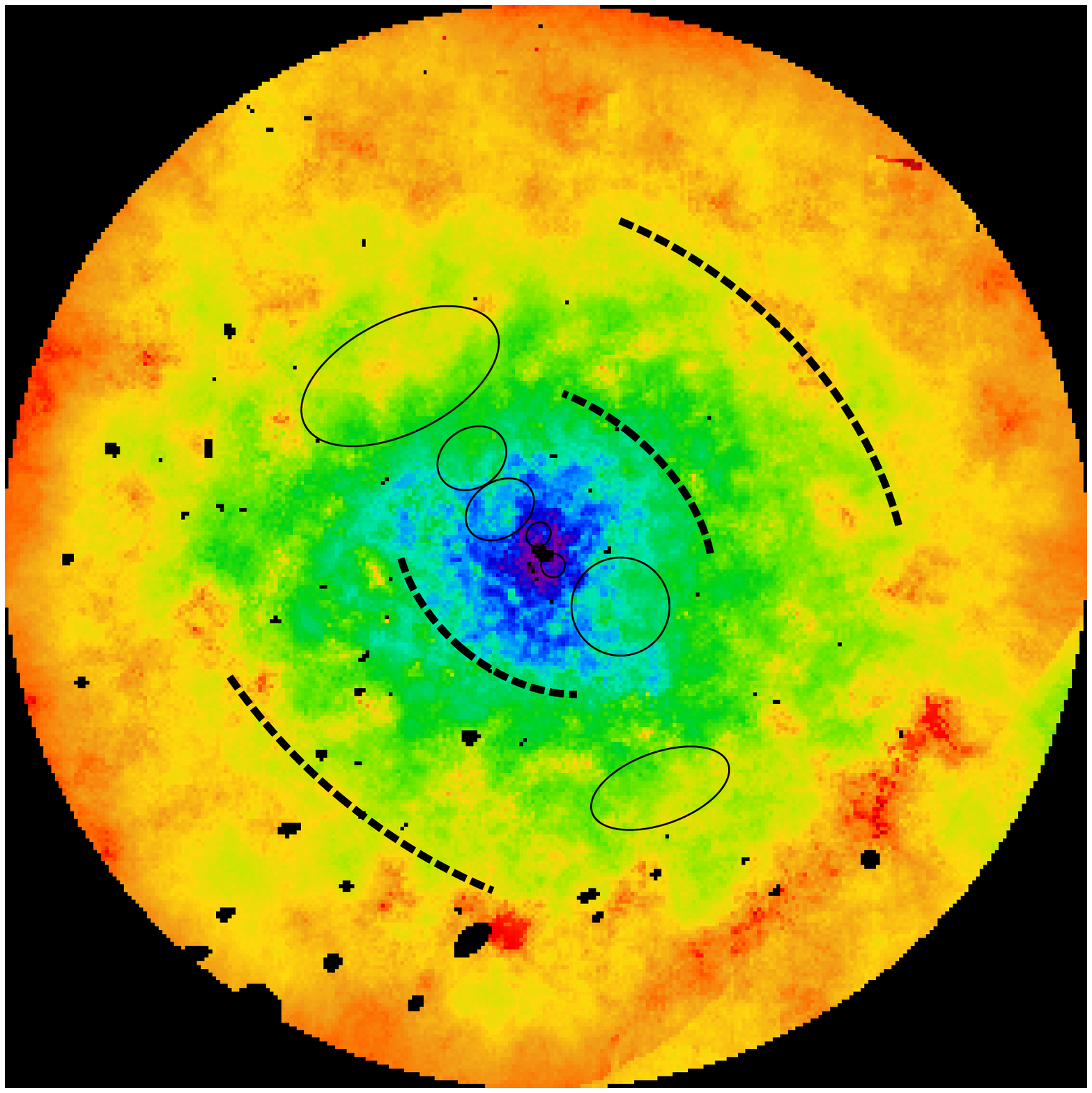}
\caption{
  Pseudo-pressure (left) and pseudo-entropy (right) maps, in arbitrary units.
  The pressure map was calculated as $kT A^{1/2}$ and the entropy map
  as $kT A^{-1/3}$, where $A$ is the {\sc apec} normalization scaled
  by the area of the extraction region.  Shocks and cavities are
  indicated as in Figure~\ref{fig:tmap}.  The pressure jumps are
  visible at the 1~kpc and 10~kpc shock fronts.  There are no
  visible entropy jumps at the shock fronts, consistent with the
  expectation that entropy jump amplitudes are small for weak shocks compared
  with those of the temperature and pressure jumps.
\label{fig:press}
}
\end{figure}

\begin{figure}
\plottwo{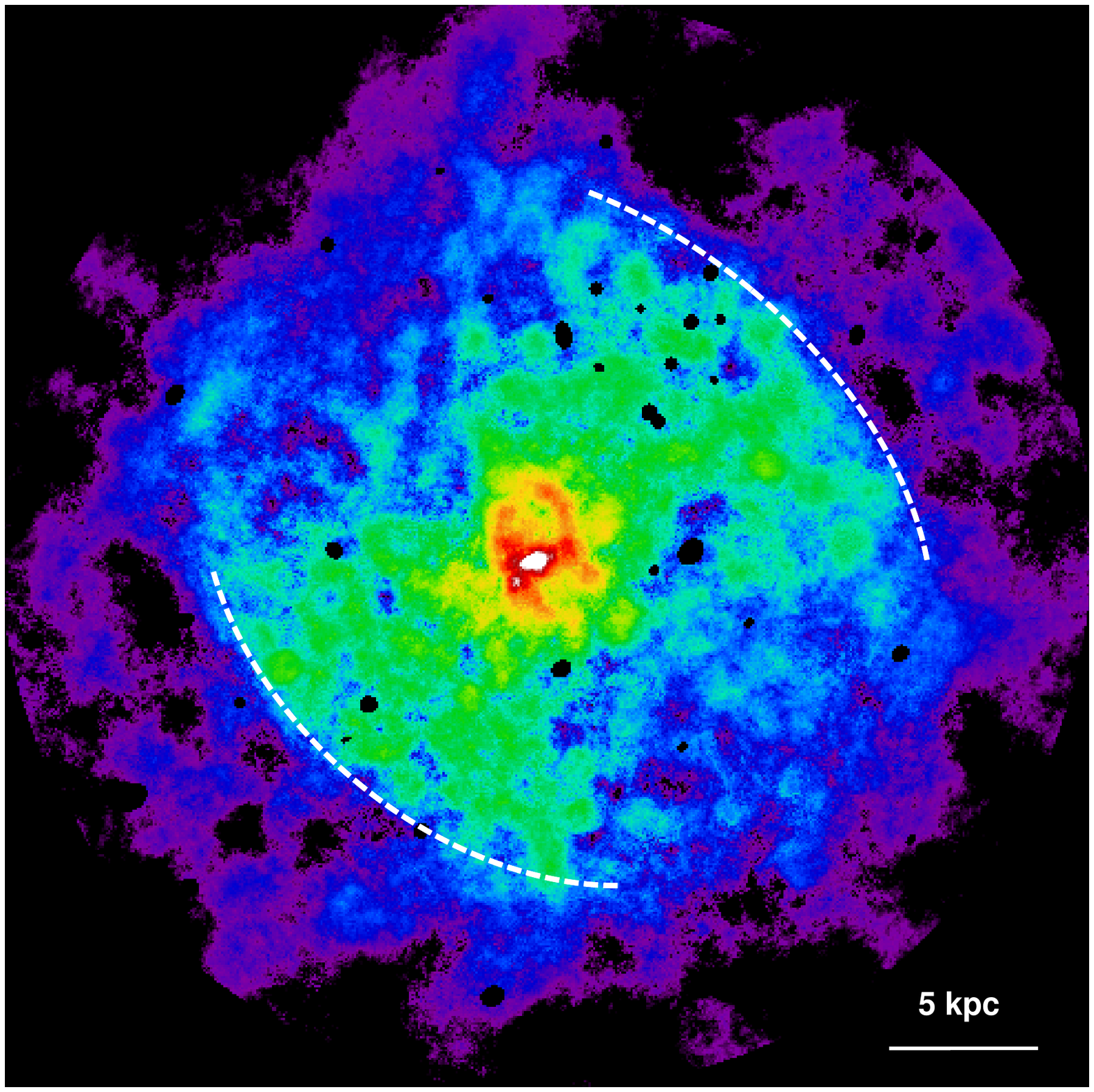}{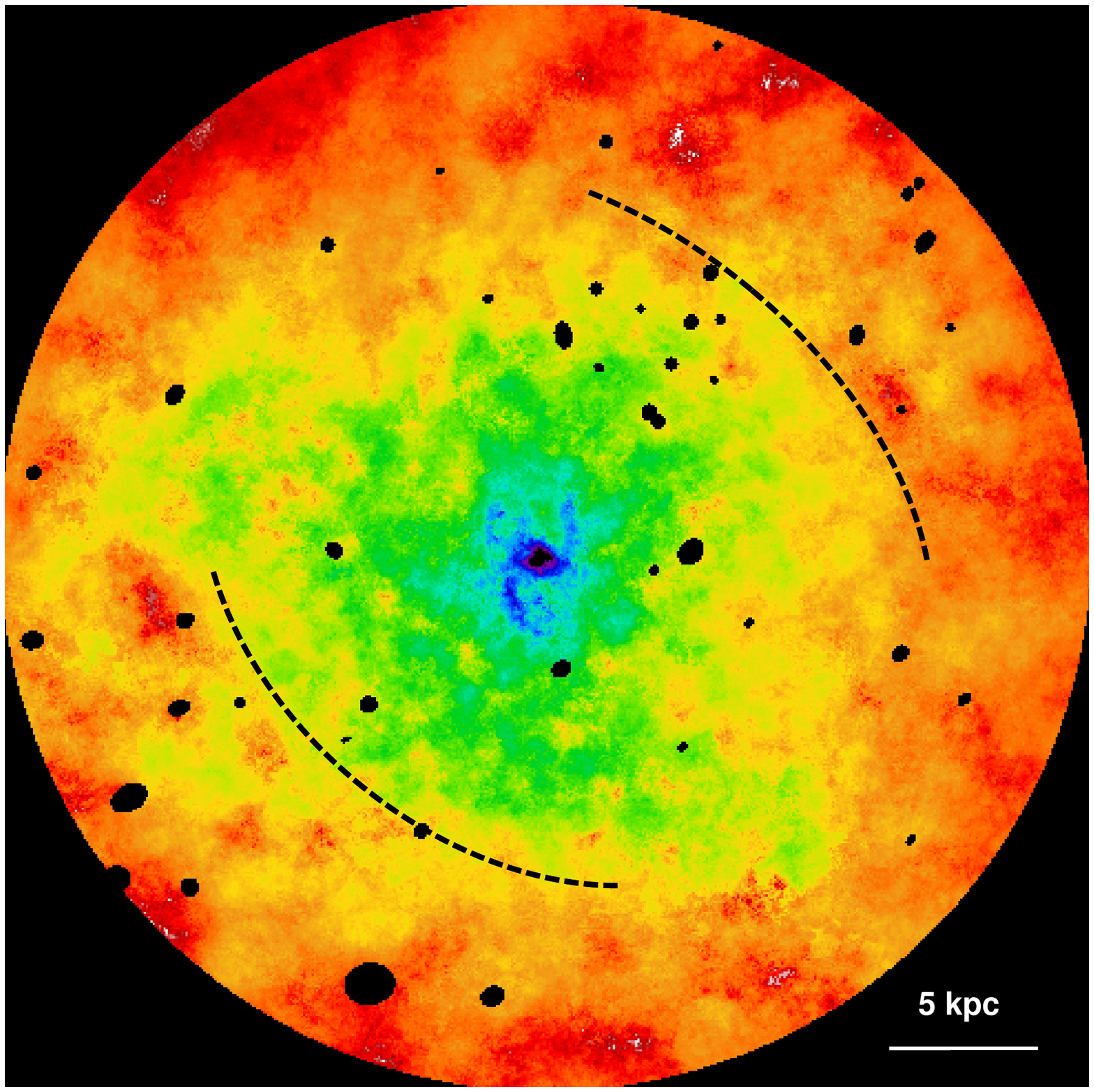}
\caption{ Pseudo-pressure (left) and pseudo-entropy (right) maps
  corresponding to the smoothed temperature map of the core shown in
  Figure~\ref{fig:core_tmap} (with the same regions overlaid), created
  as in Figure~\ref{fig:press}.  Pressure increases are clearly seen
  at the 10~kpc shock fronts and in the bright, shock heated rims
  surrounding the inner cavity pair.  There is some structure visible
  in the core of the pseudo-entropy map (seen in blue
    as filament-like structures), likely resulting from central gas
  that has been pushed out and uplifted from the core by the expanding
  and rising X-ray cavities.
\label{fig:core_press}
}
\end{figure}

We derived pseudo-pressure and pseudo-entropy maps from the smoothed
temperature maps.  The fitted {\sc apec} normalization for each
spectral map pixel is proportional to the volume integral of the
square of the electron density along the line of sight.  The square
root of the normalization therefore gives an average ``projected''
density along the line of sight.  We define the pseudo-pressure and
pseudo-entropy as $kT A^{1/2}$ and $kT A^{-1/3}$, respectively, where
$kT$ is the fitted projected temperature and $A$ is the normalization
scaled by the area of the extraction region.  The wide field and core
maps are shown in Figures~\ref{fig:press}~and~\ref{fig:core_press}.
The pseudo-pressure maps reveal prominent pressure jumps across the
10~kpc shock fronts and increased pressure in the bright central rims
(\ie, the 1~kpc shock).  There is no obvious equivalent structure in
the pseudo-entropy maps.  This is not unexpected for weak shocks.  For
the strongest shock we detect, with $M = 1.78$, the Rankine-Hugoniot
shock jump conditions for a $\gamma = 5/3$ gas give a pressure jump
at the shock front of a factor of 3.7, but a jump in
  the entropy index of only a factor of 1.12.  Thus, entropy jumps
are expected to be relatively difficult to detect, particularly in
projection.

The core pseudo-entropy map (Figure~\ref{fig:core_press}) shows some
filamentary structure in the center, on the scale of the bright shocked rims.
However, unlike the features in the pseudo-pressure map, these features do not
directly trace the bright rims.  Furthermore, the entropy is lower in
these filaments as compared with the surrounding regions, in contrast
to the small increase in entropy that is expected at the shock fronts.
These structures are likely formed as the central low-entropy gas
is pushed and pulled away from the center as the inner cavities expand
and rise buoyantly.

\begin{figure}
\plottwo{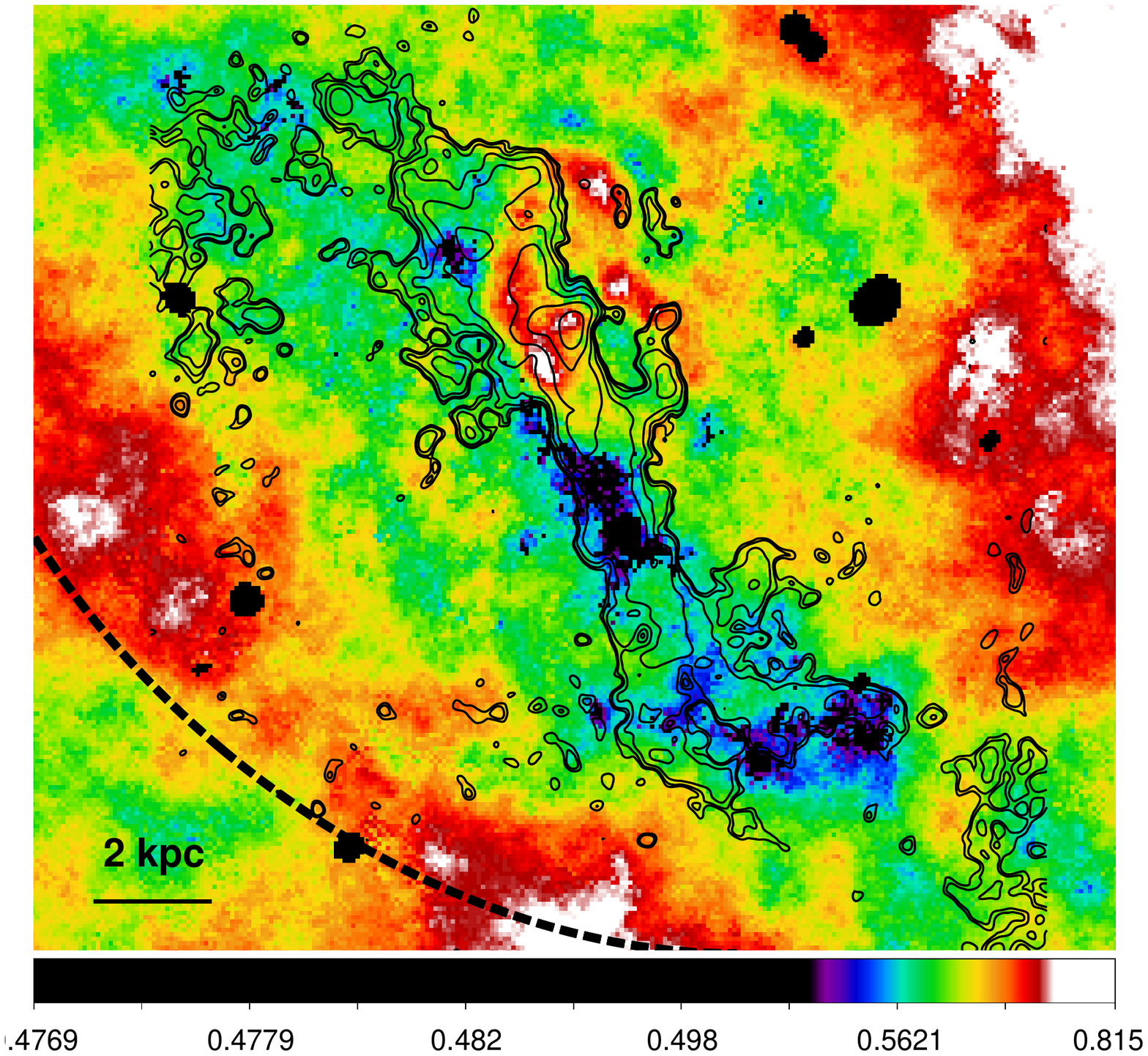}{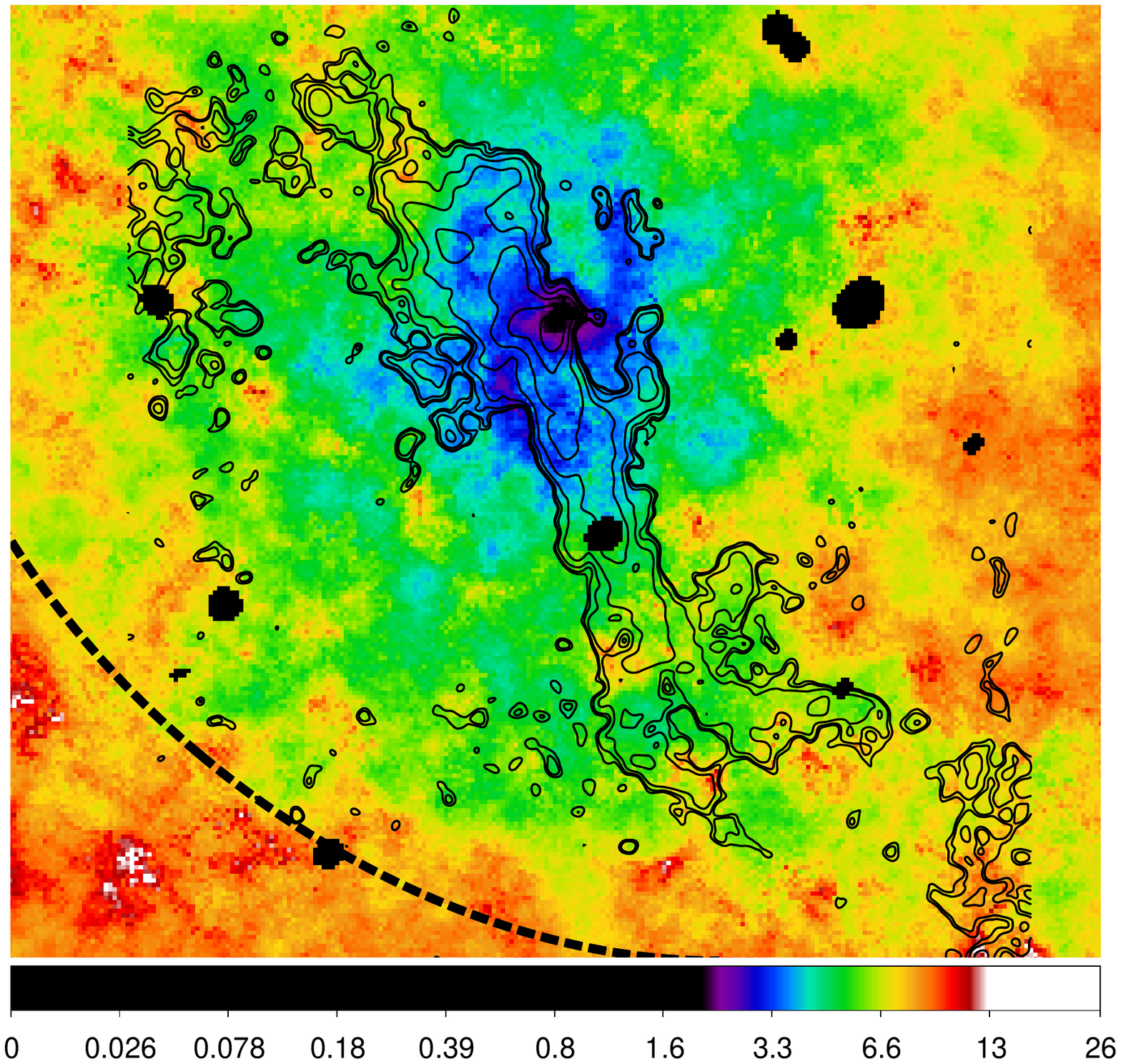}
\caption{Core temperature (left) and pseudo-entropy (right) maps with
  H$\alpha$ contours from R11 overlaid.  Units are keV for temperature
  and arbitrary for entropy.  The SE 10~kpc shock is indicated by the
  dashed line.  The H$\alpha$ emission follows the cool plume of gas up
  to the intermediate cavities, and traces the SW inner cavity.  There
  is a correlation with the detailed, filamentary structure of the central
  low-entropy gas in the pseudo-entropy map (e.g., with the
  low-entropy filament just NE of the core).
\label{fig:halpha}
}
\end{figure}

The central regions of the core temperature and pseudo-entropy maps
are shown with the H$\alpha$ contours from R11 overlaid in
Figure~\ref{fig:halpha}.  As shown in R11, the H$\alpha$ emission
follows the plume of cool X-ray emitting gas up to the location of the
intermediate cavities, indicating that these cavities have uplifted
cold phase gas from the core as they buoyantly rise.  The H$\alpha$
emission also clearly traces the SW inner cavity, suggesting that the
central cold-phase gas is dynamically coupled to the cavity and is
being pushed out as the cavity expands.  
The SE 1~kpc shock appears to be passing through the H$\alpha$
filament (at least in projection) without destroying it.
Finally, there is a
correlation between the H$\alpha$ emission and the detailed,
small-scale, filamentary structure
of the central low-entropy gas.  This is consistent with the above
suggestion that these entropy features likely arise from gas that has
been displaced from the core by the cavities, rather than from entropy
changes driven by the central shocks.

\begin{figure}
\plotone{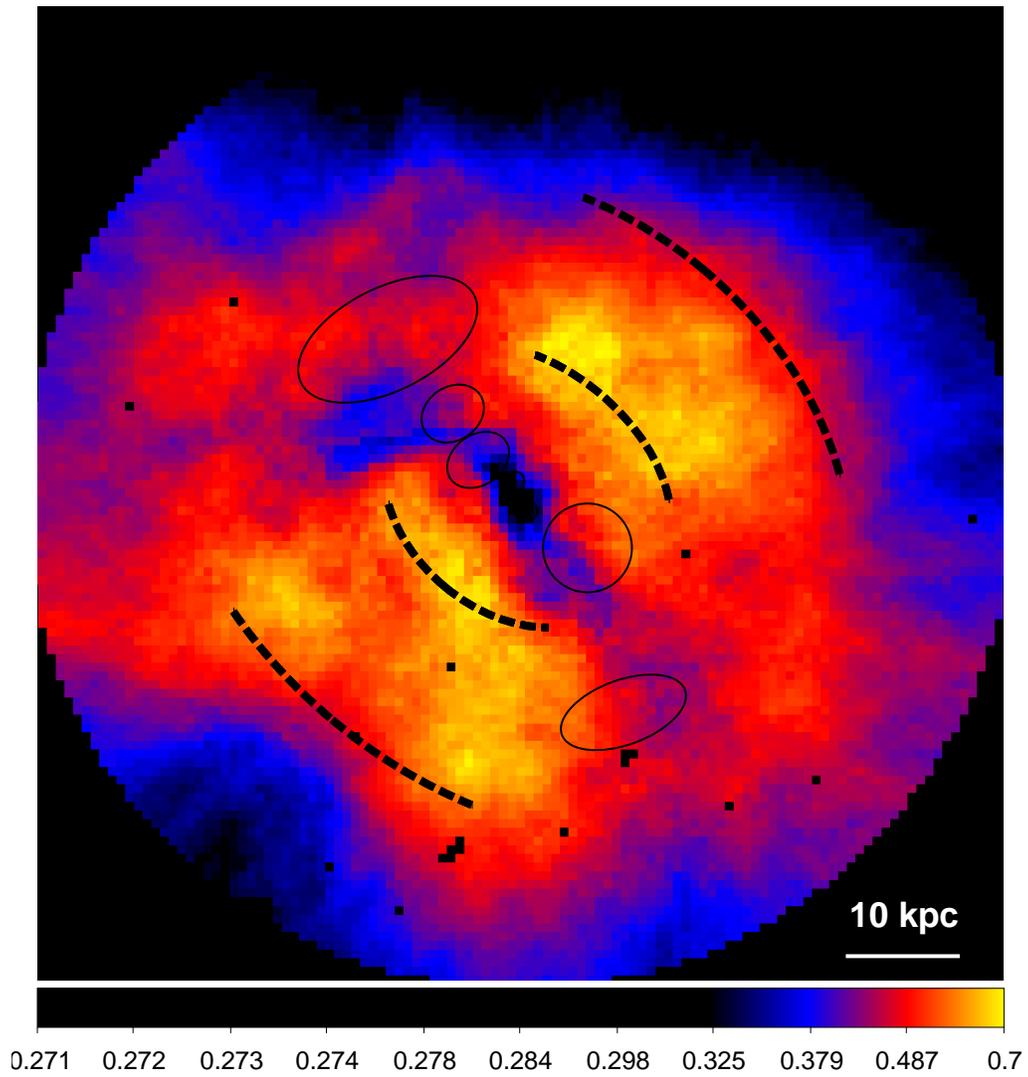}
\caption{
Smoothed abundance map, with cavity and shock regions overlaid as in
Figure~\ref{fig:tmap}.  The apparent low abundance in the plume of 
uplifted cool gas (extending NE and SW of the core) is an artifact of
fitting a single temperature model to multi-temperature spectra (see text).
\label{fig:abund}
}
\end{figure}

Since each extraction region in the above maps contained only 1500 net
counts, the abundance values determined by these fits
were not tightly constrained, and the corresponding
abundance maps show little to no structure.  Abundance measurements
in $\sim$0.7~keV gas are challenging due to the large number of
closely spaced emission lines, and the resulting degeneracy between
line and continuum emission at {\it Chandra's} energy resolution.  We
found that, for N5813, achieving 1-$\sigma$ errors on the order of
5--10\% for the abundance required roughly 30,000 net counts.
Fortunately, the new deeper observations provided well over one
million photons appropriate for use in spectral fitting, allowing us
to map the abundance distribution in the gas.  A smoothed abundance
map, with 30,000 net counts per extraction region, is shown in
Figure~\ref{fig:abund}.

The abundance map shows an apparent plume of low abundance gas that corresponds
to the plume of uplifted cool gas seen in Figure~\ref{fig:tmap}
(roughly 40~kpc long, extending from SW to NE).  It
has long been recognized that fitting multi-temperature spectra with a
single thermal model can give anomalously low abundance values (the
so-called ``Fe-bias'' effect, Buote 1999).  To test for this effect,
we fit multi-temperature models to the apparent central minimum and
apparent low-metallicity NE plume.  We find that the plume is well
described by a two-temperature model, with $kT_{\rm low} \approx
0.40$~keV, $kT_{\rm high} \approx 0.75$~keV, and an abundance (set
equal for each temperature component) of $Z_{\rm plume} \approx 45$\%
solar, consistent with the surrounding ICM.  
In contrast, we find that,
while including multiple temperature components for the central region
(in a $2.2 \times 3.8$~kpc ellipse oriented NE to SW, along the plume)
does raise the fitted abundance, it is still significantly lower than
the abundance of the surrounding ICM, at $Z_{\rm center} \approx 37$\%
solar.  
Thus, we conclude that the apparent low abundance in the extended
plume is a fitting artifact arising from the Fe-bias effect, and that
the measured abundance in the plume is consistent with that of the
surrounding ICM, while the apparent abundance minimum within the
central few kpc can not be easily explained by this effect.
We note that central abundance dips that apparently cannot be explained
solely by projection effects have been seen in other systems (e.g., Blanton
\etal\ 2003;  Panagoulia \etal\ 2013).  In addition to the apparent
low abundance plume, there 
appear to be increases in the
projected abundance at the location of the shock fronts, in
contradiction with what is expected from the Fe-bias effect.  A more
detailed study of these features and the ICM abundance in general
in N5813 will be presented in an upcoming paper.

\subsection{Radial Profiles} \label{sec:profiles}

\subsubsection{Azimuthally Averaged Profiles} \label{sec:azprofs}

\begin{figure}
\plotone{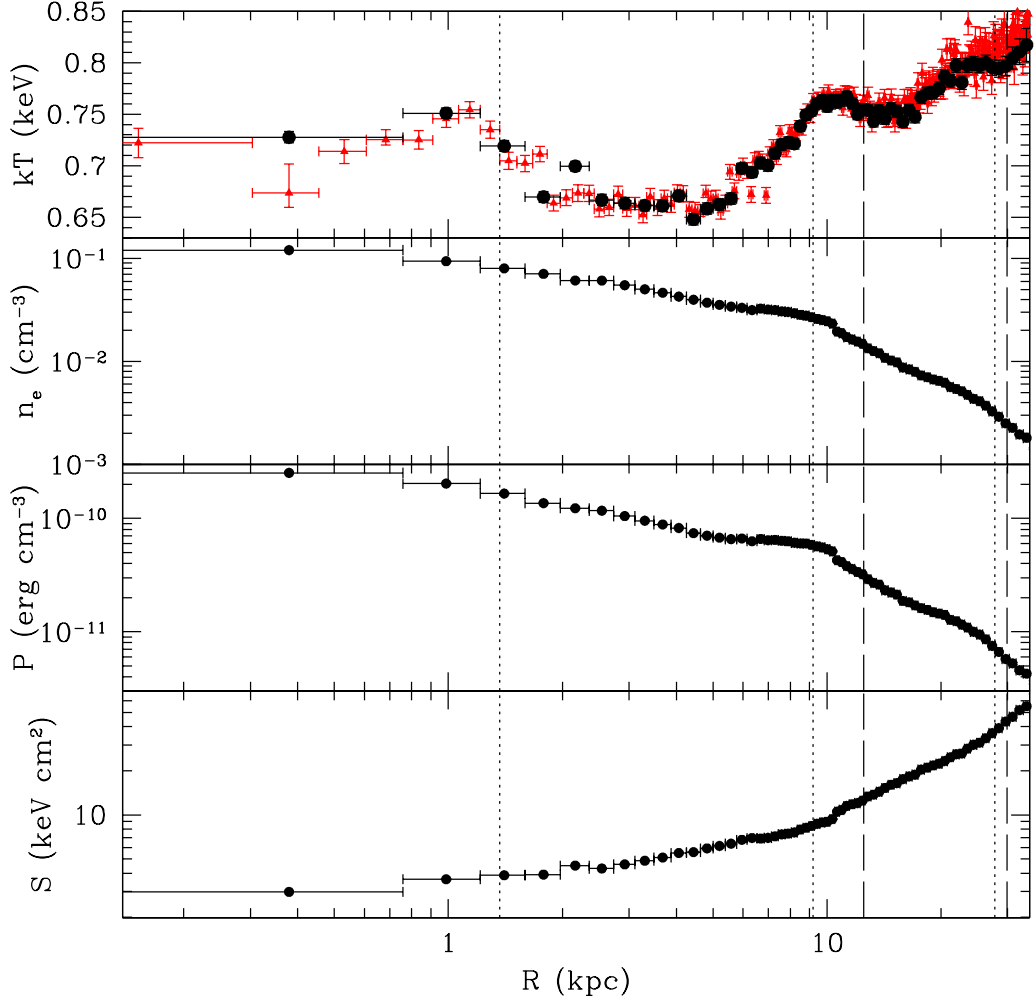}
\caption{
  Azimuthally averaged radial profiles for (top to bottom)
  temperature, electron density, pressure, and entropy, extracted from
  circular annuli.  The top
  panel shows the projected (black circles) and deprojected (red
  triangles) temperature profiles.  All other panels show deprojected
  values.  The vertical lines mark the positions of the shock fronts
  to the SE (dotted) and NW (dashed) determined by fits to the surface
  brightness profiles in sectors (see \S~\ref{sec:shock_structure}).
  Increases in temperature associated with each pair of shock fronts
  (\ie, each full elliptical edge) are clearly seen even in the
  azimuthally averaged, projected temperature profile.  The ``kink''
  in the deprojected profiles is likely due to the breakdown of the
  assumption of spherical symmetry at the location of the bright,
  sharp shock edges at $\sim10$~kpc.
\label{fig:azprof}
}
\end{figure}

We produced projected radial profiles by fitting spectra extracted
from concentric annuli, centered on the centroid of the diffuse emission
at large radii.  Each annulus was chosen to be at least 1\arcsec\
wide and contain at least 1000 net counts in the 0.6--3.0~keV band.
The inner annuli were limited by the size constraint (and therefore
contained more than 1000 net counts), whereas the outer annuli were
wider and contained roughly 1000 net counts.  The resulting temperature
profile is shown in the top panel of Figure~\ref{fig:azprof}.  The
positions of the shock fronts (determined in
\S~\ref{sec:shock_structure}) are marked by vertical lines.  Even in
the projected temperature profile in circular annuli (which
effectively smooths out the elliptical shock fronts), temperature
enhancements associated with each shock are clearly visible.
The central 1.5\arcsec\ radius region, which contains the central AGN, has
been excluded from the fit to the central region.
The central temperature rise is due to the presence of the inner shock
rims surrounding the inner cavities.

As in R11, we determined the 3D structure of the ICM by performing an
``onion peeling'' deprojection analysis (Fabian \etal\ 1981) using
concentric annuli, each 
of which contained at least 8000 net counts (in projection) in the 0.6--3.0~keV band.
We fit each annulus with an absorbed {\sc apec} model, with the
abundance fixed at 50\%~solar (consistent with results
  from spectral fits to the total diffuse emission and with typical
  values in the projected abundance map shown in
  Figure~\ref{fig:abund}).  We assume spherical shells of uniform
density.  Given the rich morphology of N5813, with elliptical shock
edges, cavities, and other features, it is clear that the assumption
of spherical symmetry does not strictly apply.  Thus, the deprojected
profiles include some level of systematic uncertainty associated with
the assumed geometry.  The deprojected temperature, density, pressure,
and entropy profiles are shown in Figure~\ref{fig:azprof}.  We see
that the deprojected temperature is generally consistent with the
projected temperature.  We interpret this an indicating that the scale
of the azimuthally averaged temperature gradient is generally small
compared with the sizes of our extracted annuli, so that projection
effects in any given annulus (which are dominated by emission from
adjacent annuli) are relatively small.  The small discontinuous jumps,
or ``kinks'',
in the density, pressure, and entropy profiles around 11~kpc are
associated with the sharp surface brightness edges to the NW and SE
from the intermediate shock.

\subsubsection{Shock Front Profiles} \label{sec:shprofs}

To confirm the temperature rises across the surface brightness edges seen in
Figures~\ref{fig:tmap}~and~\ref{fig:core_tmap}, we derived temperature
profiles in annular bin sectors across each of the edges.  The
profiles were centered on the center of curvature for each feature
(such that there was a bin boundary that traced the surface brightness
edge), then the distances were corrected to measure the distance from
the center adopted for the azimuthal profiles presented in
\S~\ref{sec:azprofs}.

\begin{figure}
\plottwo{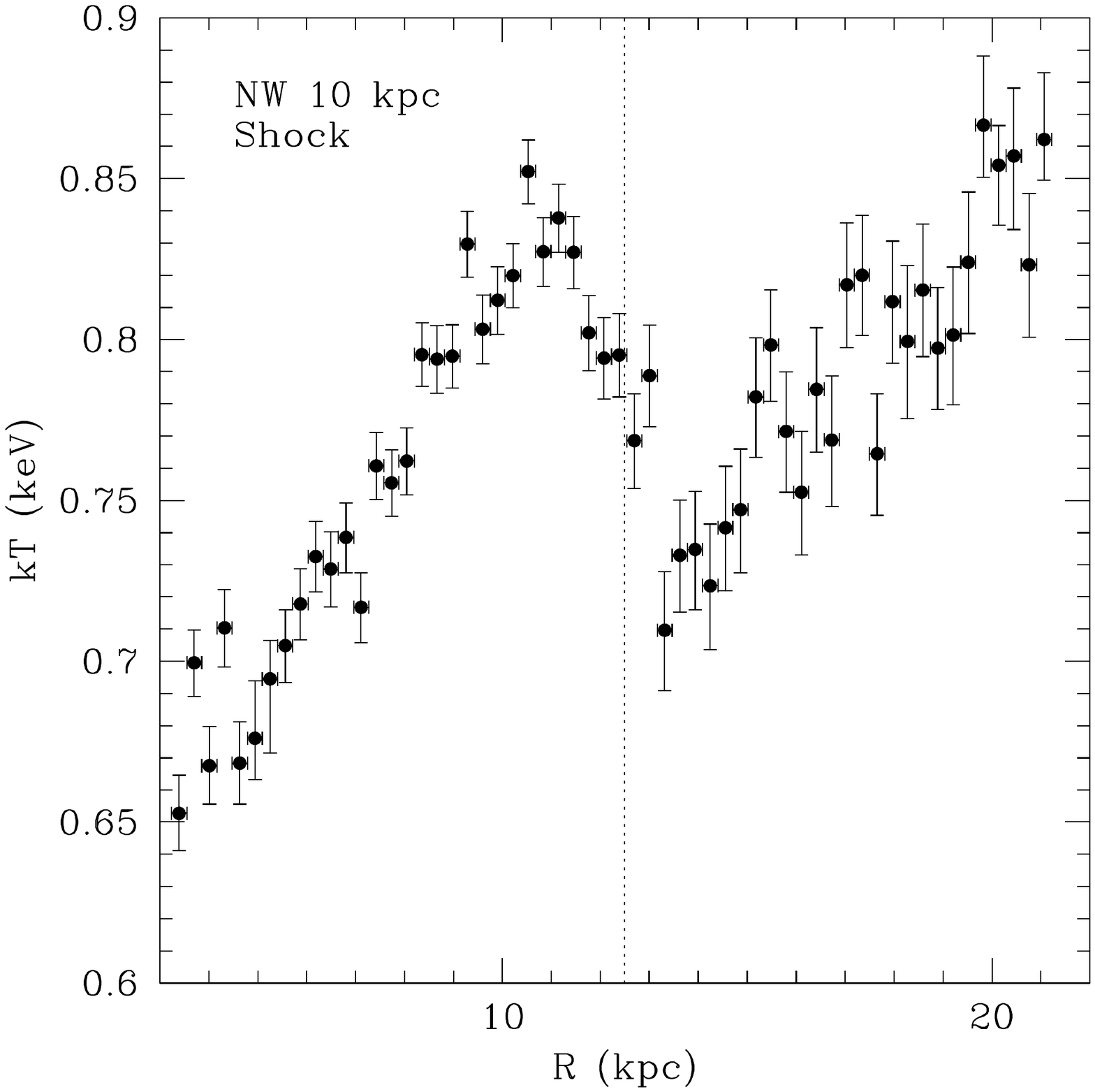}{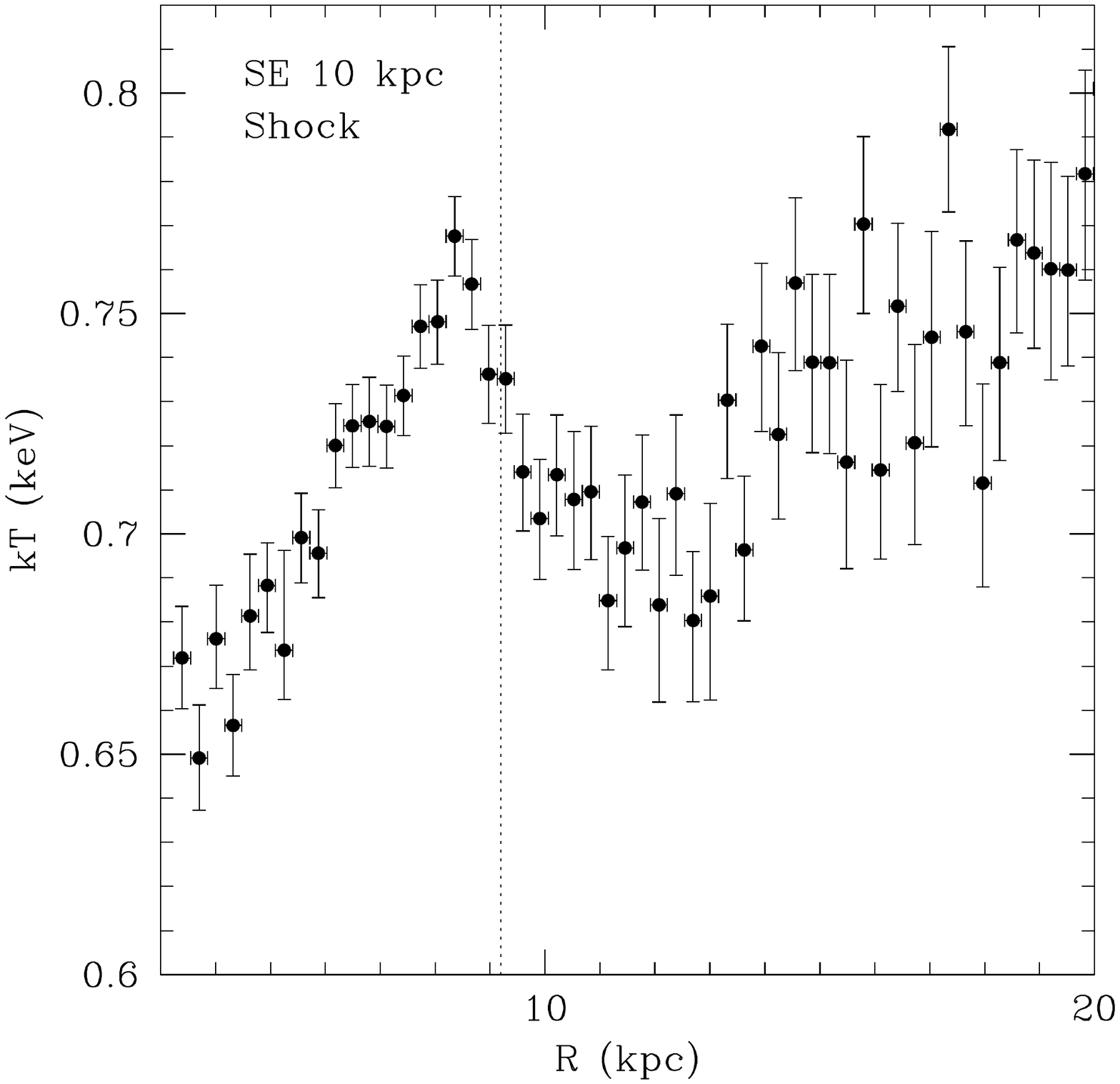}
\caption{
Projected temperature profiles across the NW (left) and SE (right)
10~kpc shock fronts.  The dashed lines indicated the fitted shock
front positions.
\label{fig:midsh_ktprofs}
}
\end{figure}

\begin{figure}
\plottwo{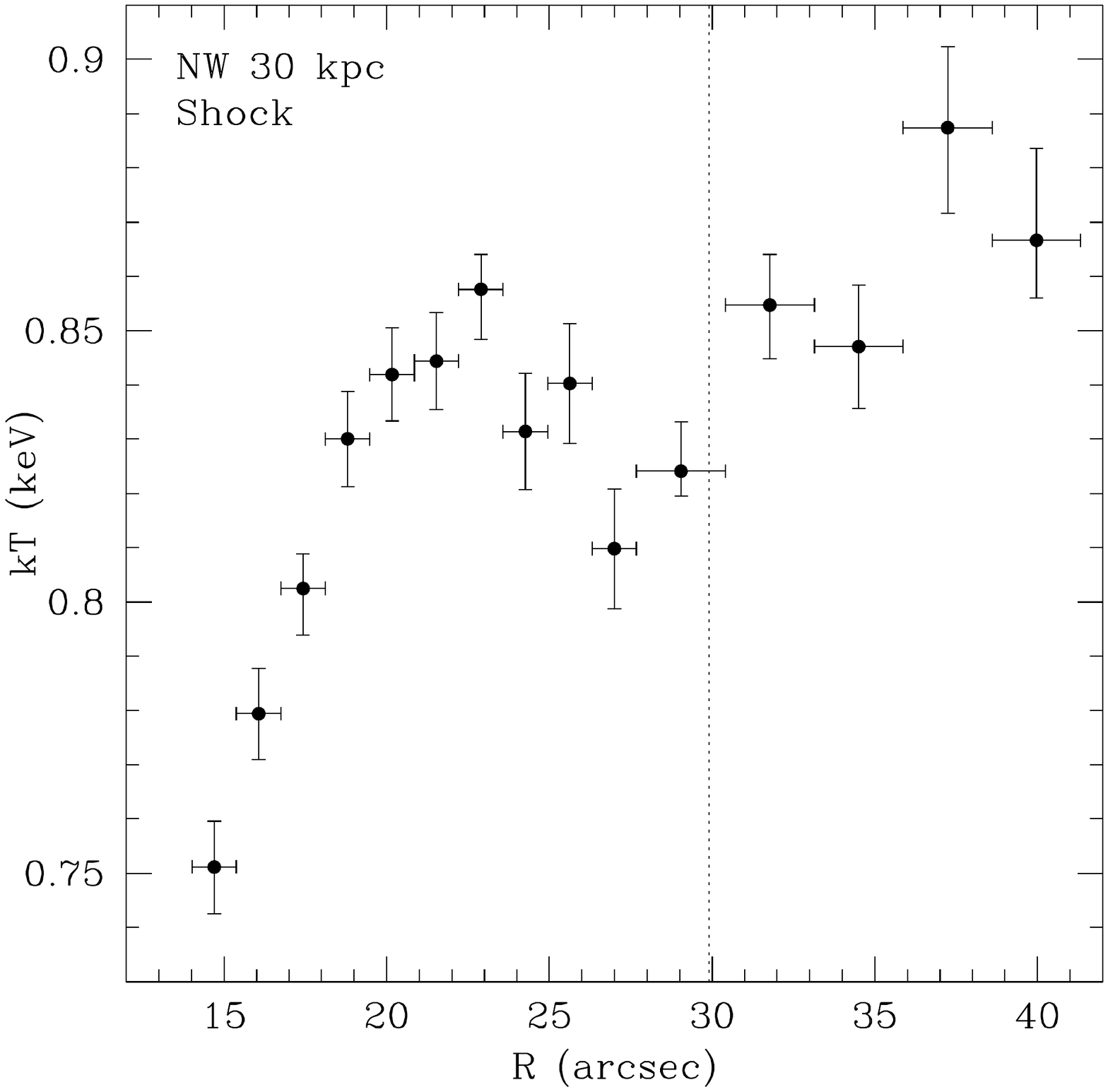}{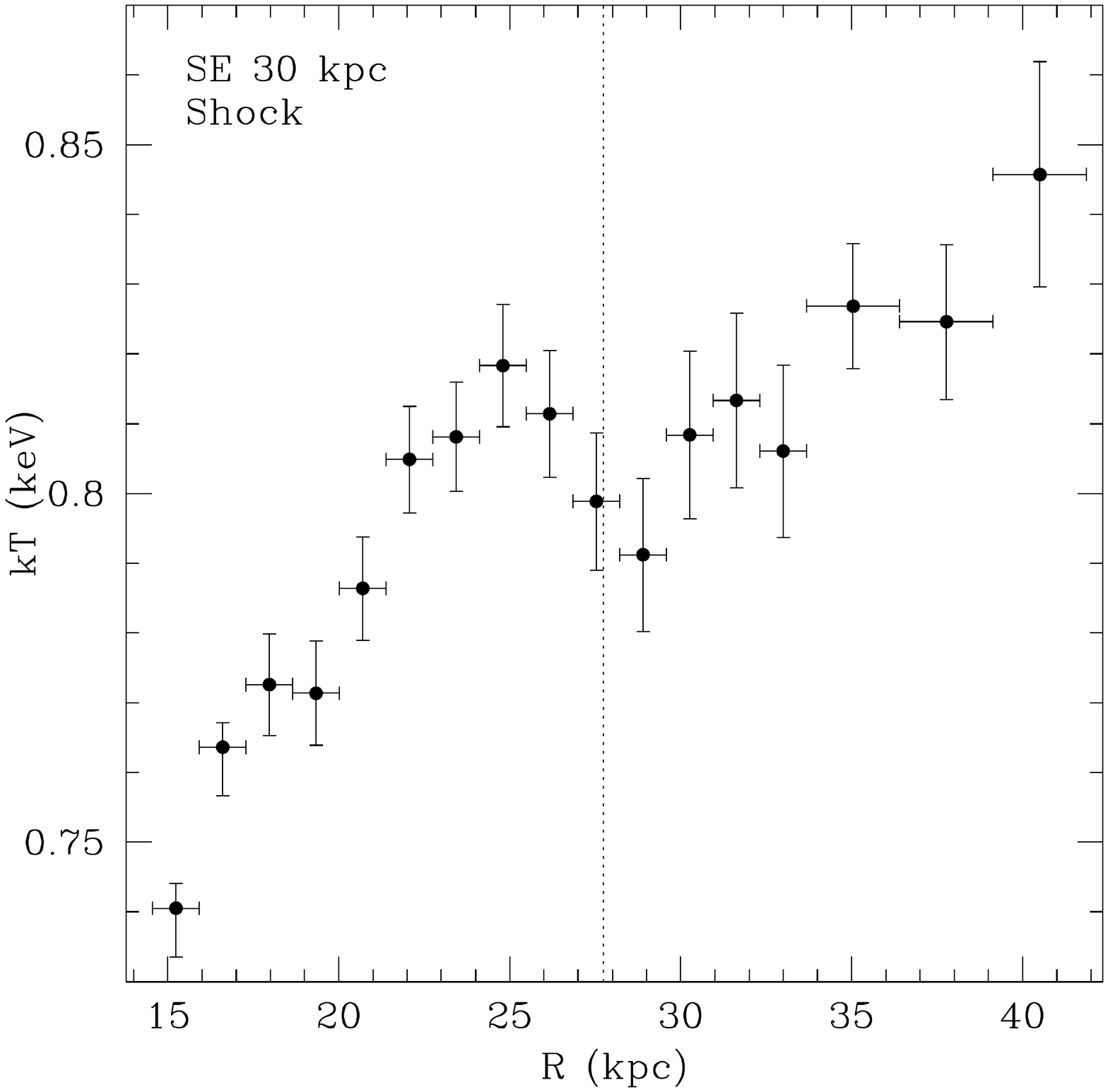}
\caption{
Projected temperature profiles across the NW (left) and SE (right)
30~kpc shock fronts.  The dashed lines indicated the fitted shock
front positions.
\label{fig:outsh_ktprofs}
}
\end{figure}

The temperature profiles across the 10~kpc and 30~kpc shock fronts are
shown in Figures~\ref{fig:midsh_ktprofs}~and~\ref{fig:outsh_ktprofs},
respectively, with the fitted shock front positions overlaid (see
\S~\ref{sec:shock_structure}).  Clear temperature rises are seen at
each front, confirming that these features are shocks.  In the case of
the 10~kpc shocks, the deeper observations allow us to measure the
temperature profiles with greater angular resolution (on the scale of
a few hundred pc) as compared with R11.  In addition to the overall
shift in normalization due to the updated {\sc AtomDB} (see
\S~\ref{sec:spec}), the peak change in temperature across the fronts
is larger as compared with the more coarsely binned profiles in R11.
 In contrast, the shock Mach numbers that we
derive here (\S~\ref{sec:shock_structure}) are consistent with those
derived in R11, since the surface brightness profile requires fewer
counts per radial bin, allowing for smaller bins.
This is consistent with our suggestion in R11 that the temperature
jumps we found there were smaller than what was expected based on the
Mach numbers derived from the surface brightness jumps due to
projection effects. 

In some cases, there appears to be shock heated gas just {\it outside}
the positions of the shock front (\eg, the two bins to the right of
the edge in Figure~\ref{fig:midsh_ktprofs},
left).  There are at least two factors that can contribute to this effect.
First, there is a small uncertainty in the fitted locations of the
shock fronts.  Second, we show in \S~\ref{sec:shock_structure} that
the shock fronts have finite widths (on the order of 0.5~kpc for the
10~kpc shock).  When fitted with a discontinuous model, the preferred
position of the edge will naturally be at the center of the smoothed front.

\section{Structure of the Shock Fronts} \label{sec:shock_structure}

\begin{figure}
\plotone{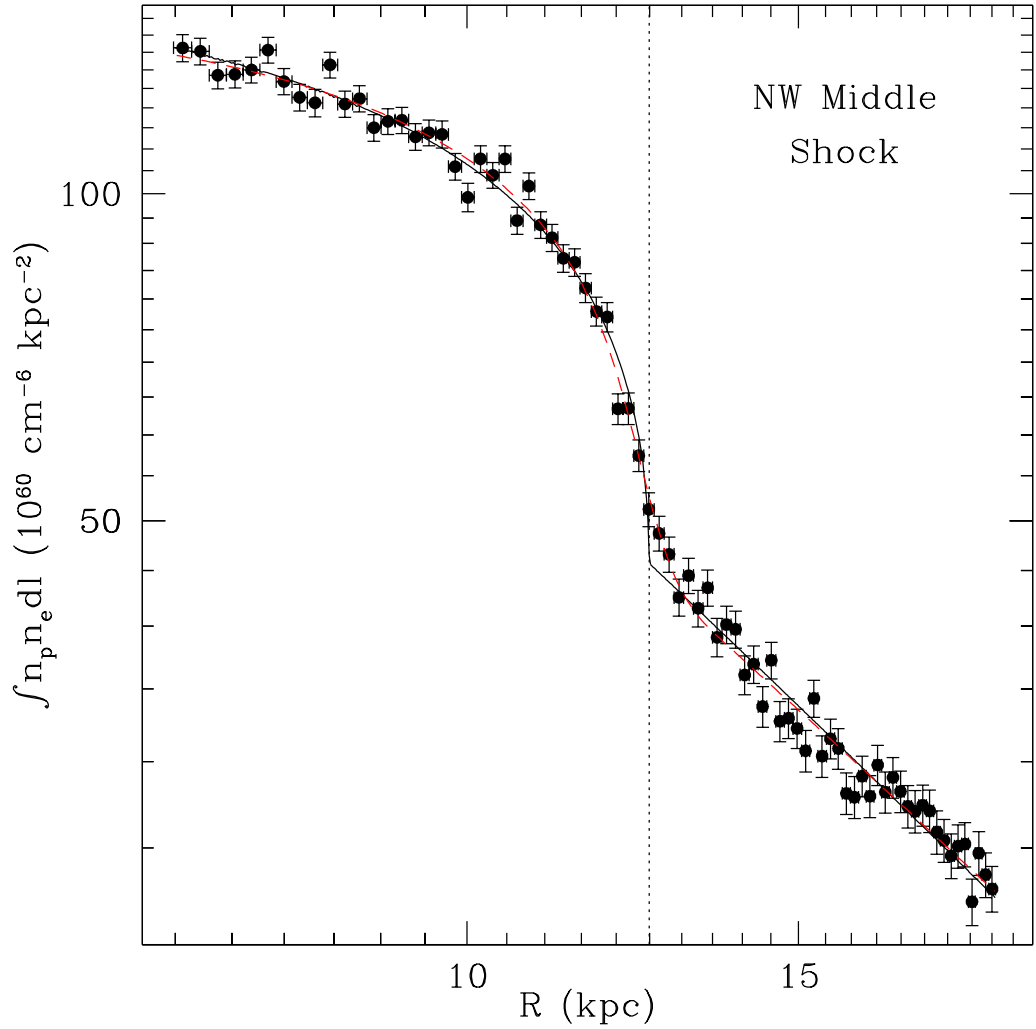}
\caption{Integrated emission measure profile across the NW middle
  ($\sim10$~kpc) shock.  The black solid line shows the fitted
  projected model
  consisting of a discontinuous power-law density profile (with
  separate inner and outer power-law slopes).  The red dashed line
  shows the fit with an identical model smoothed with a Gaussian, with
  the Gaussian
  width allowed to vary.  The vertical dotted line shows the location
  of the shock front.  The Gaussian smoothed model gives a
  better fit to the data.
\label{fig:nwedge}
}
\end{figure}

To quantitatively characterize the structure of the shock fronts, we
applied the edge fitting method described in R11.  We
extracted the 0.3-3.0~keV surface brightness profiles across each of the three
shocks (at roughly 1~kpc, 10~kpc, and 30~kpc) to the SE and the NW, where the
edges are sharpest.  We excluded the NW side of the 1~kpc shock as
this region shows a complicated morphology, with the shocks driven by
the NE and SW inner cavities just beginning to overlap
(Figure~\ref{fig:core_ximg}), and is thus
not well described by our simple spherically symmetric model.  
Although the 1~kpc SE shock edge appears relatively regular, we cannot
rule out the possibility of some systematic bias in our results due to
overlapping cavity rims.  The level of this potential bias is
difficult to characterize without detailed numerical simulations,
although we do not expect it to strongly affect our results.
Each surface brightness profile was centered on the center
of curvature for the corresponding edge, which was not coincident with the
location of the AGN or with the adopted center for the azimuthal profiles
derived in \S~\ref{sec:azprofs}.  These surface brightness profiles were
converted into integrated emissivity profiles (IEM) using the best
fitting projected temperature in each bin (determined from fitting a
single absorbed {\sc apec} model) with the abundance fixed at 50\%
solar.  The IEM profiles were then fit by projecting a spherically
symmetric discontinuous
power-law density model.  The free parameters of this model are the
normalization, the inner density slope, the outer density slope, and
the amplitude and location of the density jump.  Finally, the
density jumps were converted into shock Mach numbers using the
standard  Rankine-Hugoniot shock jump 
conditions for a $\gamma = 5/3$ gas.  The fit to the NW
10~kpc shock is shown in Figure~\ref{fig:nwedge}.  The results of all
of the fits are summarized in Table~\ref{tab:shocks} (note that all radii
are scaled to give the distance from the center adopted for the
azimuthal profiles presented in \S~\ref{sec:azprofs}, rather than the
distance from the center of curvature of each front).
We find Mach numbers of roughly 1.8, 1.5, and 1.2 for the 1~kpc,
10~kpc, and 30~kpc shocks, respectively.
These results confirm the presence of the outer shock at $\sim30$~kpc,
only hinted at in the shallower
observations analyzed in R11.  This shock is associated with the same AGN
outburst that inflated the outermost cavity pair $\sim 50$~Myr ago.

The IEM profiles of both the NW and SE 10~kpc shock fronts show edges that
are less sharp than the discontinuous power-law model.  We
therefore fit each 
edge with the same model but smoothed with a Gaussian, with the width
of the Gaussian ($\sigma$) as an additional free parameter of the fit.
The fit for the NW 10~kpc shock front is shown along with the
discontinuous model fit in Figure~\ref{fig:nwedge}.
The smoothed model is a significantly better fit to the data for both
the SE and NW 10~kpc shock fronts, with both probabilities $<0.04$\%
based on the F test.
For the SE 1~kpc shock, the smoothed model is a marginally better fit,
with a probability of $\sim 9$\%, while for the 30~kpc shock
both models describe the data equally as well.
The 1~kpc SE and both 10~kpc
shock fronts are shown to have finite widths at greater than 90\%
confidence.  Only upper limits can be placed on the widths of the
30~kpc shock fronts,
where the surface brightness is relatively low.  

The shock front edges are
expected to be blurred by {\it Chandra's} PSF, which varies across the field.
Table~\ref{tab:shocks} gives the size of the 50\% encircled energy
fraction (EEF) region at the location of each shock, assuming a 0.74~keV plasma
with an abundance of 50\% solar and Galactic absorption, and the
fitted edge widths.  The 90\% EEF is about twice as wide as the 50\% EEF for
this model.  For the SE 1~kpc shock front, the edge width is on the order of
the size of the PSF.  Thus, the measured width of this edge may not be
intrinsic to the front.  However, for the NW and SE 10~kpc
shock fronts, the 90\% confidence lower limit on $\sigma$ is larger than the
local PSF. The difference is particularly significant for the NW
10~kpc shock, where the PSF is smaller (the optical axis is located NW
of the central core).  Furthermore, we note that the best fit shock
width is somewhat larger (consistent within the errors) for the NW
10~kpc shock front, as compared with the 
SE 10~kpc front, despite the fact that the PSF is more than twice as
large in this 
region.  This shows that the measured widths are not driven by
PSF blurring.  

We conclude that, in the case of the 10~kpc shock, the edge widths
given in Table~\ref{tab:shocks} are likely intrinsic to the edges
themselves and not instrumental artifacts.  In the
case of the 30~kpc shock, only upper limits on the shock
widths can be placed.  
The intrinsic widths may simply be due to irregularities in the shape
of the shock front, due to variations in the shock strength
with azimuthal angle created as the shock propagates through a non-uniform
and/or turbulent ICM.  To test this effect, we fit the surface brightness
profiles, as above, in sectors of varying angular width for the SE and
NW 10~kpc shock fronts, excluding very wide sectors that include the X-ray
cavities.  We found consistent density jumps for every case that we
tried, although for very narrow sectors the uncertainties were large.  Thus,
while we cannot rule out the possibility that the finite shock front
widths that we find are due to deformations in the shock fronts
(possibly at smaller angular scales than we are able to measure), we
do find that determining the shock widths in more narrow sectors does not
result in significantly smaller widths.
Possible implications of finite shock widths and the impact of a
turbulent ICM are discussed further in
\S~\ref{sec:transport}.

\section{Discussion} \label{sec:discuss}


\subsection{Heating, Cooling, and Feedback in the ICM} \label{sec:feedback}

Shocks are expected to heat the ICM as they propagate.  While there is
a relatively large temperature jump at the shock front, this increase
is transient as the gas subsequently expands and cools.  The relevant
quantity for lasting heating of the gas is the change in entropy.
The equivalent amount of heat energy $\Delta Q$
imparted to the gas by a shock due to a 
change in entropy $\Delta S $ is given by
\begin{equation} \label{eq:dq}
\Delta Q \simeq T \Delta S = E \Delta \ln \frac{p}{\rho^\gamma}, 
\end{equation}
where $E = C_{\rm V} \, T$ is the total thermal energy of the gas,
$\gamma$ is the adiabatic index (taken to be 5/3), and $p$ and
$\rho$ are the pressure and density, respectively.  Thus, each shock contributes a fraction
$\Delta \ln \frac{p}{\rho^\gamma}$ of the total thermal energy
in the gas.  In the case of weak shocks, this fraction is small
 (between 0.4--12\% for the shocks we detect in N5813).
To
compare with radiative cooling, we are interested in the cumulative
effect of shock heating per local cooling time in the gas (\ie, the
number of shocks per local cooling time and the total amount of heat
energy deposited in the gas).  We approximate the local cooling time
just outside each shock edge as the time it would take the gas to
radiate away all of its thermal energy based on its
  deprojected emissivity.
Based on the shock ages given in
Table~\ref{tab:shocks}, and keeping in mind that these ages are
expected to be underestimates for young shocks and overestimates for
older shocks, we assume an outburst repetition rate of one every 20~Myr.

The results from the above comparison of the shock heating and
radiative cooling rates are summarized in Table~\ref{tab:heating}.
Given the various statistical and systematic uncertainties associated
with, \eg, the shock Mach numbers, the shock ages and outburst
repetition rate, the deprojected cooling luminosity, etc., these
numbers should be considered to be rough estimates.   Nevertheless,
there is good agreement between the number of shocks expected per
cooling time (column 3) and the number required to completely offset
radiative cooling (column 4).  We conclude that shock heating alone is
sufficient to balance radiative cooling in the gas out to
at least $\sim$30~kpc.  This extends our earlier result in R11, where we
found a balance between heating and cooling out to $\sim$10~kpc.  At
30~kpc, the cooling time ($\sim2$~Gyr) is approaching the expected age of
the system, indicating that there is no need for additional heating
beyond this radius to completely balance cooling and explain the lack
of a strong cooling flow in this system.

We can estimate the average cavity power $P_{\rm cav}$ 
by dividing the 
total enthalpy of the cavities (estimated as $4pV$ for
  each cavity, for a total of $5.7 \times 10^{57}$~\ergs)
by the age of the oldest cavity pair ($\sim10^8$~yr; see R11, Table~3).  
This gives $P_{\rm cav} \approx 1.8 \times 10^{42}$~\ergs.
Spectral fitting gives a total X-ray luminosity of $L_X \approx 5.5
\times 10^{41}$~\ergs\ within a radius enclosing all of the cavities
($\sim26$~kpc).  Thus, in N5813 we find that a few times the $pV$
work required to inflate the cavities
is in principle large enough to offset
radiative cooling, as has generally been found in other systems
 (\eg, B\^{i}rzan \etal\ 2004; Rafferty
\etal\ 2006; Nulsen \etal\ 2007; Hlavacek-Larrondo \etal\ 2012).

Since we have shown that, in N5813, shock heating alone is sufficient
to balance cooling, one might ask whether the total heating from
cavities and shocks leads to an ``overheating'' of the ICM, breaking
the balance between heating and cooling that is required for AGN
feedback to operate over long time scales. 
  Although cavities are expected to deposit some of their enthalpy in
  the ICM as they rise buoyantly (McNamara \& Nulsen 2007), this
  heating is not directly observed, and the details on how and where
  cavities heat the ICM are poorly understood.  Furthermore, cavities
  can only directly heat the ICM locally, as they rise radially,
  whereas roughly isotropic heating must take place close to the
  central AGN to regulate feedback.  Mixing may help to distribute
  locally heated gas (\eg, Zhuravleva \etal\ 2014), although the gas
  motions must be small enough to not completely destroy the observed
  X-ray cavities.  Shocks, in contrast, heat the gas roughly
  isotropically and most strongly at smaller radii, and the thermal
  effect of shocks on the ICM is, in some cases, directly observed.
  We suggest that AGN outburst shocks may generally play a significant
  role in AGN feedback, particularly at small radii where the Mach
  numbers are higher.  The cavities are then free to rise buoyantly,
  and release their energy to heat the ICM at larger radii.  The fact
  that cavities are observed at larger radii demonstrates that they do
  not necessarily release their enthalpy to the ICM at small radii,
  consistent with this picture.  Weak shocks are generally difficult
  to detect, since the shock fronts are thin and easily masked by
  projection effects.  They are more easily detected in N5813 due to
  its proximity, low gas temperature, and regular morphology.  The
  fact that one of the handful of examples of confirmed outburst
  shocks known in clusters, in Abell~2052 (Blanton \etal\ 2011), was
  confirmed by a measured temperature jump only after very deep {\it
    Chandra} observations and careful deprojection is consistent with
  this scenario.  Similarly, no temperature jump is found associated
  with the presumed outburst shock in the Perseus cluster using
  extremely deep {\it Chandra} observations, even though it is the
  X-ray brightest cluster (Graham \etal\ 2008).  

\subsection{AGN Outburst History} \label{sec:history}

Measurements of the cavities and shocks allow us to estimate the total
mechanical outburst energy for each of the three outbursts in N5813.
Since the mechanical luminosity dominates the total AGN energy output
in N5813 (R11), as is generally the case for kinetic mode AGN feedback
(Fabian 2012), this gives us information on the total outburst energy
history of the AGN.  We take the cavity internal
  energy to be $3pV$, where the pressure $p$ is taken from the
azimuthally averaged pressure profile.  The assumed cavity dimensions
and derived internal energies are given in Table~\ref{tab:cavities}.
We use the
  cavity internal energy rather than the enthalpy (which is the sum of
  the internal energy and the work required to inflate the cavity, and
  roughly equal to $4pV$ for cavities filled with relativistic
  particles) since the work done during cavity expansion goes into
  driving shocks, the energy of which we account for separately.  As
in R11, we estimate the outburst energy in shocks as
\begin{equation} \label{eq:eshock}
E_{\rm s} \approx p_1 V_{\rm s} \left( p_2/p_1 - 1 \right), 
\end{equation}
where 
$p_1$ and $p_2$ are the pre- and post-shock pressures, respectively,
and $V_{\rm s}$ is the total volume enclosed by each roughly
ellipsoidal shock surface.  
Although
equation~\ref{eq:eshock} is expected to slightly underestimate the
shock energy as the Mach numbers were larger at earlier times (giving
larger pressure jumps), in R11 we showed that this estimate agrees
well with shock energies derived from simple 1D hydrodynamic
simulations of a central point explosion by matching the simulated
surface brightness profiles with the observations.  The total
mechanical energy output of each outburst is simply the sum of the
total cavity and shock energies.
 We note that the ratio of the energy in cavities to that in shocks is between
0.15--0.3 for each outburst, and that in the case of the central
outburst, where the cavities occupy a significant fraction of the
volume enclosed by the 1~kpc shock surface, this ratio is somewhat
lower than the expected value of $\sim 1$.  This is likely a result of
underestimating the central gas pressure, and hence the central cavity energies,
due to the complicated 
morphology in this region, which leads to systematic uncertainties associated with
projection effects and a lack of knowledge on the volume filling factor of the
X-ray emitting gas.  However, we note that such an underestimate
would only moderately affect our total outburst energy, well within
the factor of a few at which we expect our simple estimates to be
accurate, and hence would not affect our main conclusions.

Using the above estimates, we find total outburst energies of
 $8.5 
\times 10^{56}$~erg, $9.9 \times 10^{57}$~erg, and $8.9 \times
10^{57}$~erg for the outbursts associated with the 1~kpc, 10~kpc, and
30~kpc shocks, respectively.
As in R11, we find the outburst energy of the
1~kpc outburst to be more than a factor of ten less than that of the
10~kpc outburst. Based on this, we concluded that either the
1~kpc outburst is ongoing, with the AGN actively inflating the
cavities, or that the mean jet power varies significantly over
timescales on the order of 10~Myr.  Here,
we find remarkable agreement between the total outburst energies of
the 10~kpc and 30~kpc outbursts.  This suggests that the mean
outburst (or jet) power, averaged over the outburst interval of a few 10~Myr, is
fairly stable, and that the total energy of the 1~kpc outburst
is lower because this outburst is ongoing.  

In R11, we estimated the outburst power as the total energy divided by
the age of the outburst.  Here, we instead estimate this power as the
total energy divided by the duration of each outburst, which we take
to be the shock age for the youngest outburst (Table~\ref{tab:shocks})
and the outburst interval ($\sim20$~Myr, \S~\ref{sec:feedback}) for
the remaining outbursts.  This gives mean outburst powers of  $1.6
\times 10^{43}$~\ergs, $1.6 \times 10^{43}$~\ergs, and $1.4 \times
10^{43}$~\ergs\ for the 1~kpc, 10~kpc, and 30~kpc outbursts,
respectively.  Thus, we find the mean outburst power is indeed roughly
constant, certainly within a factor of a few, which is the level at which we
expect our rough estimates to be accurate.  We note that in R11 we
found a lower power for the 1~kpc outburst.  This difference arises
mainly from using results from our numerical model in R11, which
predicts a lower shock energy and larger shock age as compared with
the estimates based on observations alone (see Table~2 in R11). The
source of this difference is likely to be a systematic error that
arises when applying the spherically symmetric model to the 1~kpc
shock, which clearly has a more complicated morphology (see
Figures~\ref{fig:core_ximg}~\&~\ref{fig:core_tmap}).

We stress that our result on the consistency of the outburst power only
applies to the mean power averaged over the outburst interval of a few
10~Myr.  AGN luminosities are known to vary by up to several orders of
magnitude over very short (observable) timescales (\eg, Harris \etal\
2009).  However, in principle the outburst signatures in N5813 could
be created with a constant jet power.  In this case, the cavities
expand rapidly and drive shocks just after they are formed.  The
expansion rate drops as the {\it fractional} energy input rate
decreases, and the shocks separate from the cavities and propagate
outwards.  As the cavities rise buoyantly, they eventually disconnect
from the central jets (once the buoyant speed exceeds the expansion
rate), which then begin inflating a new pair of cavities and the
process repeats (similar models are discussed, \eg, in Fabian \etal\
2003).
Therefore, we use the term ``outburst'' to refer to the
creation of a cavity pair and its associated shock, rather than a
rapid increase in jet power, since these features are consistent with
a kinetic jet power that either also cycles in outbursts or is roughly constant.

We conclude that the most likely scenario is that N5813 is in a
``steady state'' kinetic feedback mode, with the AGN outbursts roughly
equally spaced in time, and each outburst depositing a similar amount
of total energy into the ICM, such that the mean AGN outburst power is
roughly constant.  
The lower total power of the 1~kpc outburst is an indication that this
outburst is ongoing, and has yet to deposit the bulk of its energy in
the ICM.  The detection of 1.4~GHz radio emission filling the inner
cavities (from young, high-energy, non-thermal particles injected by
the AGN), along with shock heated cavity rims that have not yet
separated from the cavities (as they have clearly done for the
intermediate and outer cavities), are consistent with this
interpretation (R11).  This is also consistent with results from Allen
\etal\ (2006), who find evidence for accretion flows around central
AGN that are stable over a few million years.  The fact that N5813
shows little evidence of a recent merger or other complicated ICM
``weather'', probably contributes to it being able to maintain such
steady state feedback over long timescales.

\subsection{Transport Processes in the ICM} \label{sec:transport}

Fits to the surface brightness profiles across the shock fronts reveal
that the SE and NW 10~kpc fronts have non-zero widths that cannot be
explained by {\it Chandra's} PSF (\S~\ref{sec:shock_structure}).  
Measurements of the widths of surface brightness edges in cluster
X-ray observations have been used to place constraints on transport
processes in the ICM by comparing these widths with the local collisional
mean free paths of the particles.  Edges that are significantly
narrower than the particle mean free path indicate that Coulomb
diffusion is suppressed across the edge.
In the case of cold fronts in clusters, the front widths are
found to be significantly smaller than the particle mean free paths
(\eg, Vikhlinin \etal\ 2001a; Russell \etal\ 2012), implying that
Coulomb diffusion is suppressed across the fronts.  However, in
cold fronts the suppression of transport processes (and
Kelvin--Helmholtz instabilities) is
likely due to magnetic draping, where magnetic field lines are
stretched along the cold front edge as the relatively cool, dense gas
moves through 
the ICM (Vikhlinin \etal\ 2001b; Asai \etal\ 2004,
2005, 2007; Lyutikov 2006; Dursi \& Pfrommer 2008, ZuHone \etal\
2011). In this case we consider shock fronts which propagate through the
ICM, and therefore magnetic draping is not expected to occur.

In the case of merger shocks,
Markevitch \& Vikhlinin (2007) find a bow shock width of roughly
35~kpc, on the order of the local particle mean free path,
for 1E~0657-56 (the Bullet cluster), 
although this width is only marginally preferred over a zero width front.
Russell \etal\ (2012)
examine shock widths in the merging cluster Abell~2146, which
contains a leading bow shock and a trailing reverse shock.  They find
finite width shocks in each case, with the reverse shock width being
significantly smaller, and the bow shock width being marginally
smaller, than the particle mean free path.  They conclude that
transport processes are suppressed across these merger shocks.
In the case of AGN outburst shocks, Croston \etal\ (2009) 
consider the thickness of the northeastern shock in Centaurus~A, but
their results are inconclusive.

It is therefore of interest to compare our measured shock widths with
the Coulomb mean free path of particles in the ICM.   
In the region of the shock fronts there are four relevant mean free
paths: that in the pre-shock region, $\lambda_{\rm in}$, that in the
post-shock region, $\lambda_{\rm out}$,
that of particles crossing the front from the post-shock to the
pre-shock region, $\lambda_{\rm in \rightarrow out}$, and that of particles
crossing the front from the pre-shock region to the post-shock
region, $\lambda_{\rm out \rightarrow in}$.  In our case, the largest
and most relevant mean free path is $\lambda_{\rm in \rightarrow
  out}$, which is given by
\begin{equation}
\lambda_{\rm in \rightarrow out} = 15 
\left( \frac{T_{\rm out}}{7 \, {\rm keV}} \right)^2 
\left( \frac{n_{e,out}}{10^{-3} \, {\rm cm}^{-3}} \right)^{-1}
\frac{x \, G(1)}{G(\sqrt{x})}
\; \; {\rm kpc},
\end{equation}
where $n_{e,out}$ is the pre-shock electron density, $x = T_{\rm
  in}/T_{\rm out}$, $T_{\rm in}$ is the post-shock gas temperature,
$T_{\rm out}$ is the pre-shock gas temperature, $G(y)=[\phi (y) - y
\phi' (y)]/2y^2$, and $\phi (y)$ is the error function (Spitzer 1962).
When calculating $\lambda_{\rm in \rightarrow out}$, pre-shock
temperatures were taken from the projected temperature profiles just
outside the shock fronts.  Post-shock temperatures were calculated
from the pre-shock temperatures by applying the Rankine-Hugoniot jump
conditions, rather than taken from the observed projected post-shock
temperatures.  We note that this is a conservative assumption from the
perspective of finding a mean free path that is smaller than the shock
width, as it gives a higher post-shock temperature (that is not
diminished by projection effects), which gives a larger value for
$\lambda_{\rm in \rightarrow out}$.

The relevant particle mean free path is given for each shock front in
Table~\ref{tab:shocks}.  The uncertainty in the mean free path is
dominated by the uncertainty in the pre-shock temperature, which is on
the order of 10\%, leading to a mean free path uncertainty of roughly
20\%.  In all cases, the particle mean free path is significantly
smaller than the shock width, on the order of ten times smaller than
the 90\% confidence lower limit for the 1~kpc and 10~kpc shocks (where
lower limits on the shock widths can be placed).
 For a weak shock, the shock width is expected to be
  roughly $w \approx \lambda/(M - 1)$, where $\lambda$ is the
  effective particle mean free path (Landau \& Lifshitz 1987; McNamara
  \& Nulsen 2007), or 1.4--2 times $\lambda$ for the weak shocks we
  consider here.  Thus, the apparent shock widths are too large to be
  explained by particle diffusion alone.  We note that this result
  likely precludes using these shock fronts to place accurate
  constraints on the electron-ion equilibration timescale, as has been
  done, \eg, by Russell \etal\ (2012).

As noted in \S~\ref{sec:shock_structure}, a measured finite shock width
may arise due to deformations in
and broadening of a shock front as it propagates through a ``clumpy'' and/or
turbulent ICM.  
Nulsen \etal\ (2013) 
provide an estimate
of the expected shock width due to
turbulence as a function of radius as the shock propagates through a
uniformly turbulent ICM.
The shock width ($w$, defined as the rms of the
  displacement of sections of the
shock front due to turbulence) is determined by the distance traveled ($r_{\rm
  s}$), the shock speed $v_{\rm s}$,
the coherence length of the turbulence ($\ell$), and the rms
turbulent speed ($\sigma_{\rm t}$).
Here, we can invert this relation to find the turbulent speed implied
by the observed shock width:
\begin{equation}\label{eq:vturb}
\sigma_{\rm t} \approx \frac{w \, v_{\rm s}}{\sqrt{r_{\rm s} \, \ell}}.
\end{equation}

The rms turbulent speed implied by measurements of each shock front
we consider are given in Table~\ref{tab:shocks}.  Following Nulsen
\etal\ (2013), we take the coherence length to be $\ell \sim 0.1 r$
(Rebusco \etal\ 2005).  Remarkably, the implied turbulent speed is
consistent with roughly 70~\kms\ in each case.
We
note that, in principle, the turbulent speeds calculated here are
upper limits on the true ICM turbulence, as this calculation does not
include the contribution of density inhomogeneities to shock
broadening, projection effects, and the effects of an
  irregular shock front geometry (which is approximated as spherically
  symmetric in sectors), all of which will act to increase the
  apparent shock width.  However, as the calculation is only
approximate, our results should not be taken as hard upper limits.

The implied turbulent velocity of roughly 70~\kms\ is
reasonable compared with results from simulations (100--300~\kms; Lau
\etal\ 2009) and observations (\eg, de Plaa et al. 2012; Sanders \&
Fabian 2013; Zhuravleva \etal\ 2014).  While this value is on the low
end of the reported
range, we note that N5813 is a relatively isolated galaxy group
(R11), which has a regular morphology and shows no signs of a recent
major merger.  Thus, the level of merger-driven turbulence is expected
to be low.
Additionally, we note that observational measurements will
naturally be biased towards detecting systems with larger turbulent
velocities.  In several cases only upper limits can be placed, with
all reported upper limits significantly larger (a few to several
100~\kms) than the $\sim70$~\kms\ detection we report here (Sanders
\etal\ 2011; Bulbul \etal\ 2012b; Sanders \& Fabian 2013).  Finally,
this value is consistent with the range of turbulent velocities of
43--140~\kms\ 
recently reported by Zhuravleva \etal\ (2014) for the (relatively low
mass) Virgo cluster.

By considering the effects of resonant scattering on \ion{Fe}{17} line
ratios, de~Plaa \etal\ (2012) constrain the turbulent velocity in
N5813 to be $140 < V_{\rm turb} < 540 \kms$, significantly larger
than our value of 70~\kms.  However, we do not view this as a serious
discrepancy for several reasons.  First, as mentioned above, the
limits provided by equation~\ref{eq:vturb} are expected to be rough
estimates.  In particular, the coherence length $\ell$ is unknown and
the only hard constraint is that $\ell < r$.  Second, as noted by the
authors themselves, the atomic data used to calculate the line ratios
in de~Plaa \etal\ (2012) suffer from significant systematic
uncertainty.  For example, by comparing results derived with the {\sc
  spex}\footnote{http://www.sron.nl/spex} code versus {\sc
  AtomDB}\footnote{http://www.atomdb.org}, they find inferred turbulent
velocities for N5813 that differ by almost a factor of two.
Finally, we note that there are plausible physical origins for the
higher level of turbulence found by de~Plaa \etal\ (2012).  For
example, they place constraints for the central $r \lesssim 5$~kpc
only, where the turbulent velocity may be larger than the total ICM
average due to the influence of the AGN, whereas the estimates given
here are averages over the entire region interior to each shock.  The
trend of the inferred turbulent speed decreasing with radius
(Table~\ref{tab:shocks}), although not statistically significant, is
consistent with this picture.  It is interesting to note that the
turbulent velocity inferred from the width of the central $r \approx
1$~kpc shock is consistent with the range given by de~Plaa \etal\
(2012), although, as noted above, the errors are large and the effects
of the larger relative PSF are unclear.  Additionally, we note that
the resonant line ratio may also be affected by bulk flows in the ICM
(\eg, the radial gas flows behind the shock fronts), possibly leading
to an over-estimate of the turbulence.

We conclude that, while the measured shock widths are too large to be
explained by Coulomb diffusion, they are consistent with arising from
the shocks propagating through a turbulent ICM.  
The implied ratio of turbulent
to thermal pressure support, estimated as the ratio of the turbulent
energy to the thermal energy (de~Plaa \etal\ 2012), is on the order of
a few percent ($\lesssim 5\%$) for a turbulent speed of 70~\kms.
For
each shock we consider, the local thermal diffusion coefficient $\nu
\approx \lambda \, c_{\rm s}$, where $c_{\rm s}$ is the sound speed,
is significantly smaller (by up to an order of magnitude) than the
implied turbulent diffusion coefficient $\eta \approx \ell \,
\sigma_{\rm t}$.  Thus, turbulent diffusion dominates the microscopic
transport in the ICM (at least in this case, and likely in other
similar-size groups, which are expected to have similar bulk ICM
properties).

\subsection{Contents of the X-ray Cavities} \label{sec:cav_cont}

Although the X-ray cavities must be close to pressure equilibrium with
their surroundings, radio observations show that generally the
pressure support from relativistic non-thermal electrons, under
the assumption of equipartition, is
insufficient to balance the thermal pressure in the ICM (B\^{i}rzan
\etal\ 2008). 
Additional pressure support might be provided by heavy non-thermal
particles, or by high temperature thermal gas, with deep X-ray
observations providing constraints on the contribution from the
latter (\eg,  Schmidt \etal\ 2002; Blanton \etal\ 2003; Sanders \&
Fabian (2007); Rafferty \etal\ (2013)).
Here we place similar limits for the low mass galaxy group
N5813. 

To place lower limits on the temperature of thermal gas in the
cavities we use a method similar to that employed in previous studies
of other systems (\eg, Sanders \& Fabian 2007; Rafferty \etal\ 2013).
 Due to the complex thermal structure of the gas,
we do not use comparison regions outside of the cavities as in Sanders
\& Fabian (2007).  
Rather, we place limits based on fits to spectra extracted
from regions at the cavity centers, where the contrast is highest and
the relative contribution from high temperature gas inside the
cavities is expected to be the largest.

Each spectrum is fit with two absorbed {\sc apec} model components.
The first component models the contribution from the total projected
emission, while the second models hotter thermal emission from within
the cavities.  To derive conservative estimates, the temperature,
abundance, and normalization of the first component are allowed to
vary.  We find that within the cavity regions this single thermal
component is sufficient to adequately model the total projected
emission (additional thermal components do not significantly improve
the fits and are not well constrained).  For the second component, the
temperature is fixed at some value, the abundance is tied to that of
the first component, and the normalization is allowed to vary.  For an
assumed cavity volume, the upper limit on the normalization gives an
upper limit on the density.  If we further assume that the cavity is
in pressure equilibrium with its surroundings, that the pressure
is equal to the average azimuthal pressure at the cavity radius
(Figure~\ref{fig:azprof}), and that the thermal gas dominates the
pressure support inside the cavities,
an upper limit on the volume filling
fraction of the gas in the cavities (for the given temperature) is
obtained.  A lower limit on the gas temperature is obtained by
increasing the fixed temperature of the hot gas until the upper limit
on the volume filling fraction is equal to unity.

We concentrate on the cavities with the largest contrast and the most
regular morphology.  Since they are closely spaced (and therefore in a
similar environment) we fit the inner cavities together, assuming
a cylindrical geometry with a radius of $r=0.408$~(0.488)~kpc and a depth
of $d=0.689$~(0.815)~kpc for the northeastern (southwestern) central cavity.
For the intermediate cavities, only the southwestern cavity is
considered (with $r=1.82$~kpc, $d=3.01$~kpc) as it has a larger
contrast and more regular shape 
compared with the northeastern intermediate cavity, which may be split
into two cavities and may be connected to the northeastern central
cavity (R11).
Similarly, we choose the northeastern outer cavity over its
southwestern counterpart due to its larger contrast and more regular
shape, and assume an 
oblate spheroidal
geometry with major axis $a=5.89$~kpc and minor axis $b=1.90$~kpc in the
plane of the sky.  We find limits on the temperature of any
volume filling thermal gas in the cavities of $>3.4$~keV, $>16$~keV, and
$>4.5$~keV for the inner, intermediate, and outer cavities, respectively
(for comparison, the limit placed by Rafferty \etal\
2013 from deep {\it Chandra} observations of the galaxy group HCG~62
is 4.3~keV).  Our most stringent constraint of $kT>16$~keV for the
intermediate cavities implies a gas density in the cavities
that is at least 20 times lower than the azimuthal average at that
radius ($n_{\rm e} < 9 \times 10^{-4}$~cm$^{-3}$).

We note that this method relies on several assumptions.  For instance,
the hot temperature component is assumed to arise solely from gas
contained within the cavities (it may be the case that hot thermal gas
supplies only part of the pressure support, while, e.g., heavy
non-thermal particles or magnetic fields make up the difference).
Additionally, the azimuthally averaged 
pressure profile, which is used to estimate the mean external pressure
at each cavity, is derived under the assumption of spherical symmetry
(\S~\ref{sec:azprofs}), which is clearly violated at some level for
this system (Figure~\ref{fig:ximg}).  Furthermore, some assumptions about
the cavity geometries (or, more specifically, their volumes) must be
made to calculate densities and pressures.  The largest source of
error is likely from the latter assumption.  We find that an
assumed uncertainty in cavity volume of 20\% leads to
an uncertainty of $\sim 10$\% on the temperature limit (a similar
result was found by Rafferty \etal\ 2013).  

\subsection{Nature of the Northern Channel} \label{sec:nchan}

\begin{figure}
\plotone{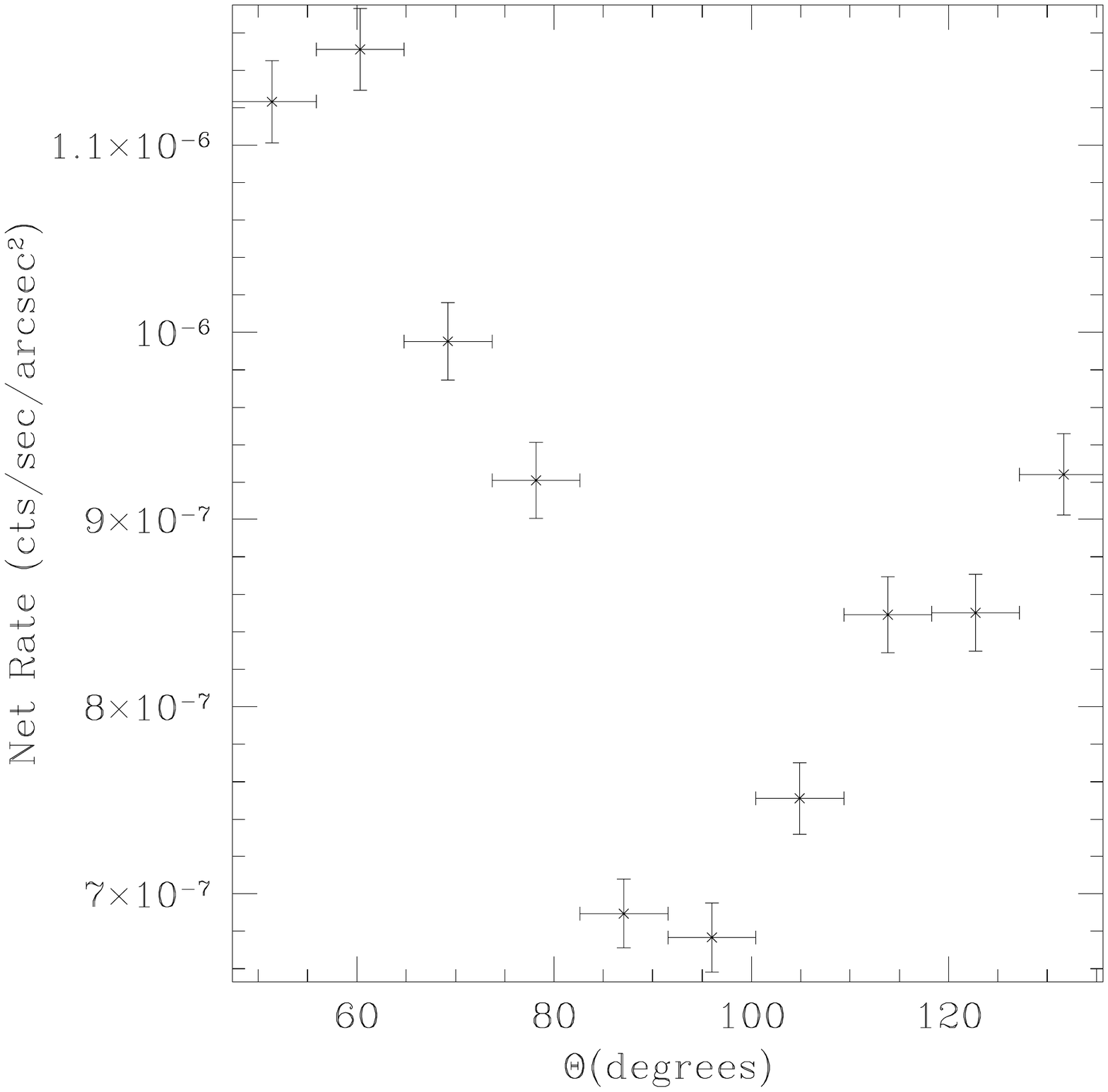}
\caption{0.3--3.0~keV surface brightness profile in azimuthal bins
  across the northern channel of decreased surface brightness shown in
  Figure~\ref{fig:ximg}.  The angle $\Theta$ is measured from east to
  north.  There is a clear dip at the location of the northern
  channel, indicating that this is a real feature.
\label{fig:nchan}
}
\end{figure}

The residual image shown in Figure~\ref{fig:ximg} reveals a
``channel'' of decreased surface brightness to the north, apparently
connected to the NE outer cavity.  To test the significance of this
feature, we extracted the surface brightness profile in azimuthal bins
roughly 9\degree\ wide and 64\arcsec\ long across the channel.  The
profile, shown in Figure~\ref{fig:nchan}, shows a clearly significant
dip across the channel, indicating that this is a real feature.

Since the northern channel appears to connect to the NE outer cavity,
one interesting possibility is that this channel is the result of energy
(presumably mostly in the
form of non-thermal particles) leaking through the bright cavity rim
and heating the ICM.  Indeed, the temperature maps shown in
Figure~\ref{fig:tmap} show relatively high temperatures in the region
of the northern channel, although the correlation between the high
temperature region in the temperature map and the channel in the residual map is not exact.  Although cavities are expected to heat the ICM
(see \S~\ref{sec:feedback}), such a direct detection of this heating has
not yet been observed.  We therefore extracted temperatures both in
(wider) azimuthal bins across the northern channel, and in identical
elliptical regions within and to either side of the channel (at
roughly the same distance from the center).  In neither case did we
find a significant temperature enhancement associated with the
channel, with temperatures of all regions agreeing within roughly
1-$\sigma$.  We conclude that, although it is possible that the
northern channel is a result of energy leaking from the NE cavity and
heating the ICM, causing the gas in this region to expand to pressure
equilibrium (and subsequently cool adiabatically) and hence the density (and surface brightness) to drop, we
find little evidence to support this interpretation.  In
particular, we cannot rule out the possibility that this feature is
simply due to large scale structure in the ICM, \eg, associated with
previous mergers or bulk motion of the group, even in this
relatively relaxed system.  
Another possibility is that this deficit is a left over cavity from an
even older, fourth outburst that lies off of the axis defined by the other
three cavity pairs, although the fact that it is at least partially
within the 30~kpc shock front is difficult to reconcile with this
interpretation.
Deep, low-frequency radio observations
would be useful to both detect the putative low-energy non-thermal
particles in the NE outer cavity and to see if this emission extends
out along the northern channel, consistent with it arising from a
leaking cavity.

\section{Summary} \label{sec:summary}

We have presented results from our analysis of a very deep {\it Chandra}
observation of the galaxy group N5813.  
This observation confirms an outer cavity pair and associated shock
(only hinted at in R11), giving a total of three pairs
of roughly collinear cavities, with each pair associated with an
elliptical shock front with a measured temperature jump.  
The derived Mach numbers are 1.8, 1.5, and 1.2 for the 1~kpc, 10~kpc,
and 30~kpc shocks, respectively.
These features are cleanly separated signatures from three distinct
outbursts of the central AGN.  
We compare the shock heating rate to the
radiative cooling rate locally at each shock front, and find that the
rates closely balance one another.  This demonstrates that shock
heating alone is sufficient to offset gas cooling and enable AGN
feedback to operate within at least the central 30~kpc (compared with
10~kpc in R11).  We suggest
that shock heating is likely important at small radii in other cool
core clusters and groups, but that in most cases the shocks are more
difficult to detect due to higher ICM temperatures, larger distances,
and more complicated ICM structure, or ``weather'', instigated by
mergers. We find that the total outburst energies of the old
and intermediate age outbursts are roughly equal, while the energy of
the young outburst is more than a factor of 10 less.  However, the
mean powers of all of the outbursts are roughly the same, within a
factor of two, indicating that the mean kinetic luminosity of the AGN
has remained stable for at least 50~Myr.  We 
suggest that the young outburst is ongoing, having deposited only a
fraction of its total energy into the ICM.  The proximity of the
central cavities to the AGN, the relatively high-frequency radio
emission that fills the cavities, and the shock heated rims
surrounding the cavities (that have not yet separated from the
cavities, as is the case for the older outbursts) are all consistent
with this scenario.
  We find that the
10~kpc (and possibly the 1~kpc) elliptical shock front is broadened by
$\sim 0.4$~kpc ($\sim 0.1$~kpc), more than ten times the particle mean
free path and thus too broad to be caused by particle diffusion.
While we cannot rule out broadening due to a clumpy ICM and/or
projection effects, using rough estimates we find that the measured
shock widths (which are upper limits for the 30~kpc shock) are all
consistent with broadening due to propagation through a turbulent ICM with a turbulent speed
of $\sim 70$~\kms.  This is within, but on the low end of, the range of
turbulent speeds expected based on simulations and other observations,
and thus provides a robust upper limit on the turbulence since other
factors may contribute to the total shock broadening.  This suggests
that transport due to turbulence dominates that of particle diffusion
throughout the ICM in N5813. 
Using spectral fits in the cavity
regions, we place lower limits on the temperature of any volume
filling gas that could completely balance the pressure within the cavity with
that in the external ICM.  Our most stringent limit of $kT > 16$~keV
comes from the intermediate cavity pair.  Finally, we find a channel of
decreased surface brightness extending north from the outer NE cavity.
We suggest that this feature may be due to energy leaking from the
cavity and heating the ICM, although we find no temperature
enhancement in the channel to support this scenario.

\section*{Acknowledgments}
Support for this work was partially provided by the Chandra X-ray Center
through NASA contract NAS8-03060, the Smithsonian Institution, and by
{\it Chandra} X-ray observatory grant GO1-12104X.  Basic research in radio astronomy at the Naval Research Laboratory is supported
by 6.1 Base funding. We thank A. Foster,
R. Smith, H. Russell, and C. Sarazin for useful discussions.

\clearpage

\begin{deluxetable}{lllc}
\tablewidth{0pt}
\tablecaption{{\it Chandra} X-ray Observations \label{tab:obs}}
\tablehead{
\colhead{Obs ID}&
\colhead{Date Obs}&
\colhead{CCDs Used}&
\colhead{Cleaned Exposure}\\
\colhead{}&
\colhead{}&
\colhead{}&
\colhead{(ks)}
}
\startdata
5907&2005 Apr 2&S1, S2, S3, I3&48.1\\
9517&2008 Jun 5&S1, S2, S3, I3&98.3\\
12951&2011 Mar 28&S1, S2, S3, I2, I3&73.7\\
12952&2011 Apr 5&S1, S2, S3, I2, I3&142.3\\
12953&2011 Apr 7&S1, S2, S3, I2, I3&31.7\\
13246&2011 Mar 30&S1, S2, S3, I2, I3&45.0\\
13247&2011 Mar 31&S1, S2, S3, I2, I3&35.7\\
13253&2011 Apr 8&S1, S2, S3, I2, I3&116.7\\
13255&2011 Apr 10&S1, S2, S3, I2, I3&43.0\\
\enddata
\end{deluxetable}

\begin{deluxetable}{lcccccccc}
\tablewidth{0pt}
\tablecaption{Properties of the Shocks\tablenotemark{a} \label{tab:shocks}}
\tablehead{
\colhead{ID}&
\colhead{$r$\tablenotemark{b}}&
\colhead{$\sigma$\tablenotemark{c}}&
\colhead{PSF\tablenotemark{d}}&
\colhead{$\lambda_{\rm in \rightarrow out}$\tablenotemark{e}}&
\colhead{$\sigma_{\rm t}$\tablenotemark{f}}&
\colhead{$\rho_2/\rho_1$\tablenotemark{g}}&
\colhead{$M$\tablenotemark{h}}&
\colhead{$t_{\rm age}$\tablenotemark{i}}\\
\colhead{}&
\colhead{(kpc)}&
\colhead{(kpc)}&
\colhead{(kpc)}&
\colhead{(pc)}&
\colhead{(\kms)}&
\colhead{}&
\colhead{}&
\colhead{($10^7$ yr)}
}
\startdata
1~kpc, SE&1.37&$0.08^{+0.04}_{-0.04}$&0.11&3.3&$150 \pm 80$&$2.06^{+0.08}_{-0.12}$&$1.78^{+0.08}_{-0.11}$&0.17\\
10~kpc, SE&9.2&$0.35^{+0.09}_{-0.08}$&0.15&25&$80 \pm 20$&$1.75^{+0.04}_{-0.03}$&$1.52^{+0.03}_{-0.03}$&1.3\\
10~kpc, NW&12.5&$0.42^{+0.12}_{-0.11}$&0.07&36&$70 \pm 20$&$1.74^{+0.03}_{-0.02}$&$1.52^{+0.03}_{-0.01}$&1.8\\
30~kpc, SE&27.7&$<1.27$&0.31&88&$< 76$&$1.26^{+0.08}_{-0.06}$&$1.17^{+0.05}_{-0.04}$&5.2\\
30~kpc, NW&29.9&$<1.03$&0.06&163&$< 62$&$1.41^{+0.12}_{-0.13}$&$1.28^{+0.08}_{-0.09}$&5.1\\
\enddata
\tablenotetext{a}{Error ranges are 90\% confidence intervals.}
\tablenotetext{b}{Distance from the adopted center ($15^{\rm h}01^{\rm m}11^{\rm s}.160$, $+1\degree42\arcmin06\arcsec.88$)
to the shock front.}
\tablenotetext{c}{Shock width, obtained by fitting a Gaussian smoothed
density jump model to the integrated emission measure profile.}
\tablenotetext{d}{Size of the {\it Chandra} PSF in the region of the
  shock edge (50\% EEF).}
\tablenotetext{e}{Mean free path of particles crossing from the post-
  to pre-shock region.}
\tablenotetext{f}{RMS speed of turbulence in the ICM as estimated from
the observed shock widths and locations.}
\tablenotetext{g}{Density jump at shock front.}
\tablenotetext{h}{Mach number.}
\tablenotetext{i}{Shock age, estimated as the travel time from the
  current position to the center point of the elliptical shock edge.
  These ages are expected to be upper limits (except for possibly the 1~kpc shock), as they assume a
  constant Mach number.}
\end{deluxetable}

\begin{deluxetable}{lcccc}
\tablewidth{0pt}
\tablecaption{Shock Heating and Radiative Cooling \label{tab:heating}}
\tablehead{
\colhead{ID}&
\colhead{$t_{\rm cool}$\tablenotemark{a}}&
\colhead{Shocks/$t_{\rm cool}$\tablenotemark{b}}&
\colhead{$\Delta Q/E$\tablenotemark{c}}\\
\colhead{}&
\colhead{($10^8$ yr)}&
\colhead{}
}
\startdata
1 kpc Shock&1.4&7&9\\
10 kpc Shock&9.2&46&21\\
30 kpc Shock&22.2&111&143\\
\enddata
\tablenotetext{a}{Local cooling time of the gas, just outside of the
  shock edge.}
\tablenotetext{b}{Number of shocks per cooling time, assuming an
  outburst repetition rate of one every 20~Myr.}
\tablenotetext{c}{Reciprocal of the fraction of the total thermal
  energy $E$ added by each shock, roughly equal to the number of shocks per
  cooling time required to offset radiative cooling.}
\end{deluxetable}

\begin{deluxetable}{lccccc}
\tablewidth{0pt}
\tablecaption{Properties of the X-ray Cavities \label{tab:cavities}}
\tablehead{
\colhead{ID}&
\colhead{$a$\tablenotemark{a}}&
\colhead{$b$\tablenotemark{b}}&
\colhead{$r$\tablenotemark{c}}&
\colhead{$E_{\rm int}$\tablenotemark{d}}\\
\colhead{}&
\colhead{(kpc)}&
\colhead{(kpc)}&
\colhead{(kpc)}&
\colhead{($10^{56}$ erg)}
}
\startdata
Inner, SW&0.95&0.95&1.3&0.5\\
Inner, NE&1.03&0.93&1.4&0.6\\
Middle, SW&3.9&3.9&7.7&13.1\\
Middle-1\tablenotemark{e}, NE&2.9&2.2&4.9&5.5\\
Middle-2\tablenotemark{e}, NE&2.8&2.4&9.3&3.8\\
Outer, SW&5.2&3.0&22.2&3.9\\
Outer, NE&8.0&4.4&18.0&15.6\\
\enddata
\tablenotetext{a}{Semi-major axis.}
\tablenotetext{b}{Semi-minor axis.}
\tablenotetext{c}{Distance from central AGN.}
\tablenotetext{d}{The internal energy of the cavity, estimated as $3 P V$.}
\tablenotetext{e}{Part of a ``split'' cavity.}
\end{deluxetable}

\clearpage

\begin{references}

\reference{}
Allen, S. W., Dunn, R. J. H., Fabian, A. C., Taylor, G. B., \& Reynolds, C. S.
2006, MNRAS, 372, 21

\reference{}
Anders, E., \& Grevesse, N. 1989, Geochim. Cosmochim. Acta, 53, 197

\reference{}
Asai, N., Fukuda, N., \& Matsumoto, R. 2004, J. Korean Astron. Soc., 37, 575

\reference{}
Asai, N., Fukuda, N., \& Matsumoto, R. 2005, Adv. Space Res., 36, 636

\reference{}
Asai, N., Fukuda, N., \& Matsumoto, R. 2007, ApJ, 663, 816

\reference{}
B\^{i}rzan, L., McNamara, B. R., Nulsen, P. E. J., Carilli, C. L., \&
Wise, M. W. 2008, ApJ, 686, 859

\reference{}
B\^{i}rzan, L., Rafferty, D. A., McNamara, B. R., Wise, M. W., \&
Nulsen, P. E. J. 2004, ApJ, 607, 800

\reference{}
Blanton, E. L., Sarazin, C. L., \& McNamara, B. R. 2003, ApJ, 585, 227

\reference{}
Blanton, E. L., Randall, S. W., Douglass, E. M., Sarazin, C. L.,
Clarke, T. E., McNamara, B. R. 2009, ApJ, 697, 95

\reference{}
Blanton, E. L., Randall, S. W., Clarke, T. E., Sarazin, C. L.,
McNamara, B. R., Douglass, E. M., \& McDonald, M. 2011, ApJ, 737, 99

\reference{}
Br\"{u}ggen, M., Hoeft, M., \& Ruszkowski, M. 2005, ApJ, 628, 153

\reference{}
Bulbul G. E., Smith R. K., Foster A., Cottam J., Loewenstein M., Mushotzky
R., \& Shafer R. 2012a, ApJ, 747, 32

\reference{}
Bulbul, G. E., Smith, R. K., Foster, A., Cottam, J., Loewenstein, M.,
Mushotzky, R., \& Shafer, R. 2012b, 747, 32

\reference{}
Buote, D. 1999, MNRAS, 309, 685

\reference{}
Churazov, E., Br\"{u}ggen, M., Kaiser, C. R., B\"{o}hringer, H., \&
Forman, W. 2001, 554, 261


\reference{}
David, L. P., Jones, C., Forman, W., Nulsen, P., Vrtilek, J.,
O'Sullivan, E., Giacintucci, S., \& Raychaudhury, Somak 2009, ApJ,
705, 624

\reference{}
David, L. P., Nulsen, P. E. J., McNamara, B. R., Forman, W., Jones,
C., Ponman, T., Robertson, B., \& Wise, M. 2001, ApJ, 557, 546

\reference{}
David, L. P., O'Sullivan, E., Jones, C., Giacintucci, S., Vrtilek, J.,
Raychaudhury, S., Nulsen, P. E. J., Forman, W., Sun, M., \& Donahue,
M. 2011, ApJ, 728, 162

\reference{}
de Plaa, J., Werner, N., Simionescu, A., Kaastra, J. S., Grange,
Y. G., \& Vink, J. 2010, A\&A, 523, 81

\reference{}
de Plaa, J., Zhuravleva, I., Werner, N., \etal\ 2012, A\&A, 539, A34

\reference{}
Dolag, K., Vazza, F., Brunetti, G., \& Tormen, G. 2005, MNRAS, 364, 753

\reference{}
Dursi, L. J. \& Pfrommer, C. 2008, ApJ, 677, 993

\reference{}
Fabian, A. C. 2012, ARA\&A, 50, 455

\reference{}
Fabian, A. C., Hu, E. M., Cowie, L. L. , \& Grindlay, J. 1981, ApJ,
248, 47

\reference{}
Fabian, A. C., Sanders, J. S., Allen, S. W., Canning, R. E. A.,
Churazov, E., Crawford, C. S., Forman, W., Gabany, J.,
Hlavacek-Larrondo, J., Johnstone, R. M., Russell, H. R., Reynolds,
C. S., Salom\'{e}, P., Taylor, G. B., \& Young, A. J. 2011, MNRAS,
418, 215

\reference{}
Fabian, A. C., Sanders, J. S., Allen, S. W., Crawford, C. S., Iwasawa,
K., Johnstone, R. M., Schmidt, R. W., Taylor, G. B. 2013, MNRAS, 344, 43

\reference{}
Fabian, A. C., Sanders, J. S., Ettori, S., Taylor, G. B., Allen,
S. W., Crawford, C. S., Iwasawa, K., Johnstone, R. M., Ogle,
P. M. 2000, MNRAS, 318, 65

\reference{}
Fabian, A. C., Sanders, J. S., Taylor, G. B., Allen, S. W., Crawford,
C. S., Johnstone, R. M., Iwasawa, K. 2006, MNRAS, 366, 417

\reference{}
Foster, A. R., Ji, L., Smith, R. K., \& Brickhouse, N. S. 2012, ApJ,
756, 128

\reference{}
Gastaldello, F., Buote, D., A., Temi, P., Brighenti, F., Mathews,
W. G., Ettori, S. 2009, ApJ, 693, 43

\reference{}
Graham, J., Fabian, A. C., \& Sanders, J. S. 2008, MNRAS, 386, 278

\reference{}
Harris, D. E., Cheung, C. C., Stawarz, L., Biretta, J. A., \& Perlman, E. S.
2009, ApJ, 699, 305

\reference{}
Heinz, S., Br\"{u}ggen, M., \& Morsony, B. 2010, ApJ, 708, 462

\reference{}
Hlavacek-Larrondo, J., Fabian, A. C., Edge, A. C., Ebeling, H.,
Sanders, J. S., Hogan, M. T., \& Taylor, G. B. 2012, MNRAS, 421, 1360

\reference{}
Kalberla, P. M. W., Burton, W. B., Hartmann, D., Arnal, E. M., Bajaja,
E., Morras, R., P\"{o}ppel, W. G. L. 2005, A\&A, 440, 775

\reference{}
Landau L. D., \& Lifshitz E. M. 1987 Course of Theoretical Physics,
Vol. 6: Fluid Mechanics (London: Pergamon)

\reference{}
Lau, E. T., Kravtsov, A. V., \& Nagai, D. 2009, ApJ, 705, 1129

\reference{}
Lyutikov, M. 2006, MNRAS, 373, 73

\reference{}
Machacek, M. E., Jerius, D., Kraft, R., Forman, W. R., Jones, C.,
Randall, S., Giacintucci, S., \& Sun, M. 2011, ApJ, 743, 15

\reference{}
Mahdavi, A., Trentham, N., \& Tully, R. B. 2005, 130, 1502

\reference{}
McNamara, B. R., \& Nulsen, P. E. J. 2007, ARA\&A, 45, 117

\reference{}
Markevitch, M., \& Vikhlinin, A. 2007, Phys. Rep., 443, 1

\reference{}
Nulsen, P. J. E., Jones, C., Forman, W. R., David, L. P., McNamara,
B. R., Rafferty, D. A., B\^{i}rzan, L., \& Wise, M. W. 2007, in Heating
versus Cooling in Galaxies and Clusters of Galaxies,
ed. H. B\"{o}hringer, G. W. Pratt, A. Finoguenov, \& P. Schuecker (Berlin:
Springer), 210 

\reference{}
Nulsen, P. E. J., Li, Z., Forman, W. R., Kraft, R. P., Lal, D. V.,
Jones, C., Zhuravleva, I., Churazov, E., Sanders, J. S., Fabian,
A. C., Johnson, R. E., \& Murray, S. S. 2013, ApJ, 775, 117

\reference{}
O'Sullivan, E., David, L. P., Vrtilek, J. M. 2014, MNRAS, 437, 730

\reference{}
 Panagoulia, E. K., Fabian, A. C., \& Sanders, J. S. 2013, MNRAS, 433, 3290


\reference{}
Peterson, J. R., \& Fabian, A. C. 2006, PhR, 427, 1

\reference{}
Peterson, J. R., Paerels, F. B. S., Kaastra, J. S., Arnaud, M.,
Reiprich, T. H., Fabian, A. C., Mushotzky, R. F., Jernigan, J. G., \&
Sakelliou, I. 2001, A\&A, 365L, 104

\reference{}
Rafferty, D. A., B\^{i}rzan, L., Nulsen, P. E. J., McNamara, B. R.,
Brandt, W. N., Wise, M. W., \& R\"{o}ttering, H. J. A. 2013, MNRAS,
428, 58

\reference{}
Rafferty, D. A., McNamara, B. R., Nulsen, P. E., \& Wise, M. W. 2006,
ApJ, 652, 216

\reference{}
Randall, S., Nulsen, P., Forman, W., Jones, C., Machacek, M., Murray,
S., \& Maughan, B. 2008, ApJ, 688, 208 

\reference{}
Randall, S. W., Forman, W. R., Giacintucci, S., Nulsen, P. E. J., Sun,
M., Churazov, E., David, L. P., Kraft, R., Donahue, M., Blanton,
E. L., Simionescu, A., \& Werner, N. 2011, ApJ, 726, 86 (R11)

\reference{}
Rebusco, P., Churazov, E., B\"{o}hringer, H., \& Forman, W. 2005, MNRAS, 359, 1041

\reference{}
Russell, H. R.,  McNamara, B. R., Sanders, J. S., Fabian, A. C.,
Nulsen, P. E. J., Canning, R. E. A., Baum, S. A., Donahue, M., Edge,
A. C., King, L. J., \& O'Dea, C. P. 2012, MNRAS, 423, 236

\reference{}
Sanders, J. 2006, MNRAS, 371, 829

\reference{}
Sanders, J. S., \& Fabian, A. C. 2007, MNRAS, 381, 1381

\reference{}
Sanders, J. S., Fabian, A. C., \& Smith, R. K. 2011, MNRAS, 410, 1797

\reference{}
Sanders, J. S., \& Fabian, A. C. 2013, MNRAS, 429, 2727

\reference{}
Schmidt, R. W., Fabian, A. C., \& Sanders, J. S. 2002, MNRAS, 337, 71

Spitzer, L. 1962, Physics of Fully Ionized Gases, 2nd ed. (New York, NY : Interscience)

\reference{}
Tonry, J. L., Dressler, A., Blakeslee, J. P, Ajhar, E. A., Fletcher,
A. B., Luppino, G. A., Metzger, M. R., \& Moore, C. B. 2001, ApJ, 546,
681

\reference{}
Vikhlinin, A., Markevitch, M., \& Murray, S. S. 2001a, 551, 160

\reference{}
Vikhlinin, A., Markevitch, M., \& Murray, S. S. 2001b, 549, L47

\reference{}
ZuHone, J. A., Markevitch, M., \& Lee, D. 2011, 743, 16

\reference{}
Zhuravleva, I., Churazov, E., Schekochihin, A. A., Allen, S. W.,
Ar\'{e}valo, P., Fabian, A. C., Forman, W. R., Sanders, J. S.,
Simionescu, A., Sunyaev, R., Vikhlinin, A., \& Werner, N. 2014,
Nature, 515, 85

\end{references}
\end{document}